\documentclass[aps,pre,floats,superscriptaddress,nofootinbib,floatfix,twocolumn,longbibliography,showkeys]{revtex4-1}
\usepackage[T1]{fontenc}
\usepackage{amssymb,amsmath,amsfonts,amsthm}
\usepackage{graphicx}% Include figure files
\usepackage{dcolumn}% Align table columns on decimal point
\usepackage{bm}% bold math
\usepackage{comment}
\usepackage{soul}
\usepackage{hyperref}% add hypertext capabilities
\usepackage{makecell}
\usepackage{changepage}
\usepackage{multirow}
\usepackage[table]{xcolor}
\usepackage[mathscr]{euscript}

\begin{document}% Force line breaks with
\title{Consequence of anisotropy on flocking: the discretized Vicsek model}
\author{Mintu Karmakar$^{\star}$}
\email{mcsmk2057@iacs.res.in}
\affiliation{School of Mathematical \& Computational Sciences, Indian Association for the Cultivation of Science, Kolkata -- 700032, India.}

\author{Swarnajit Chatterjee$^{\star}$}
\email{swarnajitchatterjee@gmail.com}
\affiliation{Center for Biophysics \& Department for Theoretical Physics, Saarland University, 66123 Saarbr{\"u}cken, Germany.}

\author{Raja Paul}
\email{raja.paul@iacs.res.in}
\affiliation{School of Mathematical \& Computational Sciences, Indian Association for the Cultivation of Science, Kolkata -- 700032, India.}

\author{Heiko Rieger}
\email{h.rieger@mx.uni-saarland.de}
\affiliation{Center for Biophysics \& Department for Theoretical Physics, Saarland University, 66123 Saarbr{\"u}cken, Germany.}
\affiliation{INM – Leibniz Institute for New Materials, Campus D2 2, 66123 Saarbrücken, Germany.}

\date{\today}

\begin{abstract}{We numerically study a discretized Vicsek model (DVM) with particles orienting in $q$ possible orientations in two dimensions. The study investigates the significance of anisotropic orientation and microscopic interaction on macroscopic behavior. The DVM is an off-lattice flocking model like the active clock model [ACM; EPL {\bf 138}, 41001 (2022)] but the dynamical rules of particle alignment and movement are inspired by the prototypical Vicsek model (VM). The DVM shows qualitatively similar properties as the ACM for intermediate noise strength where a transition from macrophase to microphase separation of the coexistence region is observed as $q$ is increased. But for small $q$ and noise strength, the liquid phase appearing in the ACM at low temperatures is replaced in the DVM by a configuration of multiple clusters with different polarizations, which does not exhibit any long-range order. We find that the dynamical rules have a profound influence on the overarching features of the flocking phase. We further identify the metastability of the ordered liquid phase subjected to a perturbation.} 
\end{abstract}


\maketitle

\def\thefootnote{{$\star$}}\footnotetext{These authors have contributed equally to this work}\def\thefootnote{\arabic{footnote}}

\section{Introduction}
\label{introduction}
Active matter systems represent a fascinating class of materials composed of self-propelled entities that convert energy into mechanical motion, leading to complex and often out-of-equilibrium behaviors~\cite{ramaswamy2010mechanics,marchetti2013hydrodynamics,gompper20202020,needleman2017active}. The emerging phenomena in active matter systems have gained significant attention in recent years due to their potential applications in various physical, biological, and engineering systems~\cite{marchetti2013hydrodynamics,vernerey2019biological,ghosh2021enzymes}. Active matter exhibits dynamic behaviors such as collective motion~\cite{marchetti2013hydrodynamics}, pattern formation~\cite{bar2020self,shaebani2020computational}, and even the ability to exhibit controlled transport~\cite{sanchez2012spontaneous}. These systems encompass a wide range of physical, chemical, and biological entities, from swimming bacteria~\cite{peruani2012collective}, mammalian herds~\cite{garcimartin2015flow}, fish schools~\cite{becco2006experimental,calovi2014swarming}, and sterling flocks to amoeba and bacteria colonies~\cite{steager2008dynamics}, to the cooperative behavior of cytoskeletal filaments and molecular motors in living cells~\cite{schaller2010polar,sumino2012large,sanchez2012spontaneous} or in vitro environments to synthetic colloidal particles equipped with motors~\cite{veigel2011moving,wong2016synthetic}. To understand and unravel the fundamental principles governing active matter systems, new models~\cite{shaebani2020computational} have been developed in the last two decades. 

The Vicsek model (VM), introduced by Vicsek and collaborators in 1995~\cite{vicsek1995novel}, provides a fundamental framework for studying the collective behavior of particles under aligning interactions. In this model, particles adjust their velocities to align with the average velocities of neighboring particles, leading to the emergence of coherent motion and ordered patterns.  VM has played a crucial role in advancing our understanding of flock dynamics~\cite{toner1995long,toner1998flocks,toner2012reanalysis,solon2015phase}. At low particle density or high noise, the particles move in random directions, and no long-range order is observed. The transition from the gas phase at high noise and low density to the polar ordered Toner-Tu phase at low noise and high density, displaying long-range order (LRO) by a coherent motion of all particles, is first order~\cite{chate2008collective}. But, in contrast to conventional first-order phase transition scenarios, the coexistence phase of the VM can manifest as either multiple bands of particles moving collectively, a phenomenon known as microphase separation~\cite{solon2015phase,solon2015pattern}, or a polar-ordered cross sea phase~\cite{kursten2020dry}, primarily driven by giant number fluctuations (GNF)~\cite{solon2015phase}.

Nevertheless, it is important to note that the VM posits a continuous range of possible directions for the motion of particles. However, when considering a scenario in which particles are constrained to discrete, equidistant angular orientations within a two-dimensional plane, such as in the active clock model (ACM)~\cite{solon2022susceptibility,chatterjee2022polar}, the VM-inspired dynamical principles governing particle alignment and movement remain uncharted territory. In a recent study on the ACM~\cite{solon2022susceptibility}, it was revealed that in large systems, any values of discrete orientations result in significant changes in phenomenology when compared to the VM. These changes include the loss of long-range correlations, the pinning of global order, and the transformation of coexistence bands into a single moving domain. Additionally, another study on the ACM~\cite{chatterjee2022polar} with different dynamical rules shows that for a small number of directions, the ACM mirrors the active Potts model (APM)~\cite{chatterjee2020flocking,mangeat2020flocking}, exhibiting macrophase separation of the coexistence region and reorientation transition of the ordered band from transverse to longitudinal motion as bias velocity is increased. Conversely, with more directions, the ACM transitions towards the VM, displaying microphase separation and transversely moving bands without the reorientation transition. Remarkably, the transition in the $q\to\infty$ limit of ACM~\cite{chatterjee2022polar}, known as the active XY model, shares the same universality class as the VM. Motivated by these findings, in this paper, we undertake an extensive computational investigation that examines in detail a $q$-state discretized version of the Vicsek model (DVM) where the rules of particle alignment and movement follow the Vicsek protocol.
 
We ask several intriguing questions that persist within the context of the DVM, e.g., (a) How does discretizing the directions of particles in the VM, affect the overall diverse collective dynamics and steady-state phases? (b) What is the impact of $q$ and system size on the coexistence region (micro- or macro-phase separation)? (c) How do the behavior of the density fluctuations, the direction of the system's global order, and the behavior of correlation functions in the liquid phase correspond with the self-organized patterns in the phase-coexistence region? (d) What is the nature of the DVM liquid phase as a function of $q$? To answer these questions, we study the DVM in an off-lattice domain, focusing on the three key factors: the anisotropy and degeneracy parameter $q$, noise level $\eta$, and system size. 

The paper is organized as follows: after introducing the model in Sec.~\ref{model}, we present our numerical results in Sec.~\ref{result}. Finally, we summarize and discuss the implications of
our findings in Sec.~\ref{discussion}.

\section{Model}
\label{model}
We consider $N$ self-propelled particles within a two-dimensional off-lattice domain of size $L_x \times L_y$ ($L_x > L_y$ for rectangular domain and $L_x=L_y=L$ for square domain) with periodic boundary conditions. Akin to the two-dimensional  VM, each point particle $i$ is endowed with an off-lattice position vector $\bm{r}_i=(x_i,y_i)$ and moves with a constant speed $v_0$ in individual directions given by a unit orientation vector ${\bm \sigma}_i=(\cos \theta_i,\sin \theta_i)$ with an orientation angle $\theta_i \in (0,2\pi)$ where 
\begin{equation}
\label{discrete}
\theta_i = \frac{2\pi n_i}{q} \, ,
\end{equation}
and $n_i = \{0, 1, 2, \cdots, (q-1)\}$ denote discrete orientations of the particles. $q$ denotes the ground state degeneracy where each particle can only have discrete orientations allowed by the $q$ value and therefore, the continuous $U(1)$ symmetry of the VM is replaced by the discrete $Z_q$ symmetry.  

At each discrete time step $\Delta t=1$, a particle $i$ with velocity $v_0$ moves a fixed distance $v_0 \Delta t$ and interacts with $\mathcal{N}_i$ neighboring particles within a circle of unit radius around it. The position evolves in the following way:
\begin{equation}
\label{r}
\bm{r}_i^{t+\Delta t}=\bm{r}_i^t+ v_0 \bm{\sigma}_i^{t+\Delta t} \Delta t \, , 
\end{equation} 
while the new orientation is determined by a projection of the updated orientation proposed by the Vicsek rule onto one of the $q$ allowed directions:
\begin{equation}
\theta_i^{t+\Delta t}= \mathbb{P}(\bar{\theta}_i^t + \eta \xi_i^t) \,
,\label{sigma} 
\end{equation} 
where $\mathbb{P}$ is the projection and $\bar{\theta}_i^t$ is the orientation angle of a spin-weighted sum
\begin{equation}
\label{sigma2}
\bar{\bm{\sigma}}_i^t = \sum_{j\in \mathcal{N}_i} \bm{\sigma}_j^t,
\end{equation}
of orientation vectors of neighboring particles. $\xi_i^t\in[-\pi,\pi]$ is a
scalar noise uniformly distributed and uncorrelated for all sites and times. Such noise is often called $white$ since it has a flat Fourier spectrum. $\eta$ is the noise amplitude. 

\begin{figure}[!t]
    \centering
    \includegraphics[width=1\columnwidth]{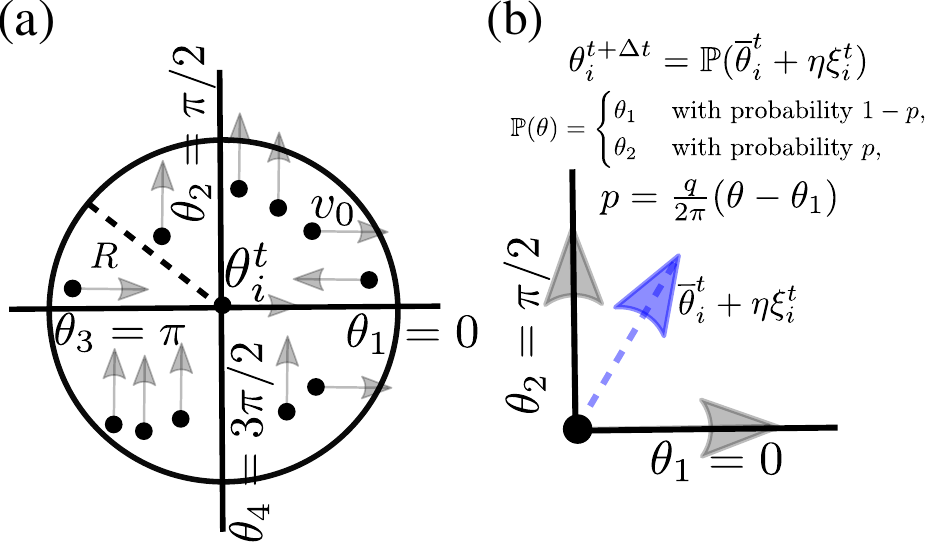}
    \caption{(Color online) Schematic of the DVM for $q=4$. (a) Four allowed orientations for the particle are $0, \pi/2, \pi,$ and $3\pi/2$ with $\theta_i^t=0$. The circular neighborhood (of radius $R$) of particle $i$ contains neighboring particles to calculate $\theta_i^{t+\Delta t}$. (b) The new orientation proposed by the Vicsek rule (blue dotted arrow) will be projected either along $\theta_1=0$ or $\theta_2=\pi/2$ probabilistically.}
    \label{schematic}
\end{figure}

We define the projection onto the allowed directions probabilistically by
\begin{equation}
\mathbb{P}(\theta) =
\begin{cases}
\theta_1 & \text{with probability} \ 1-p \, ,\\
\theta_2 & \text{with probability} \ p \, ,
\end{cases}
\end{equation}

where $\theta_1$ and $\theta_2$ are the two allowed directions which
are closest to $\theta$, such that $\theta_1=2 \pi n/q < \theta$
and $\theta_2=2 \pi (n+1)/q > \theta$ for some $n$. The probability $p\in[0,1]$ is given by $p=\frac{q}{2\pi}(\theta-\theta_1)$, i.e., minimal (0) for $\theta$ close to $\theta_1$ and maximal (1) for $\theta$ close to $\theta_2$ (cf. Fig.~\ref{schematic}). Note that then for all particles going into the direction, say, $\theta_1$, the probability to turn stochastically into another direction, say $\theta_2$, is of the order $\sim$ $\frac{q}{2}\eta$, i.e. small for small $q$ and noise $\eta$. 

Notice that the dynamical rules governing DVM are different from the $q$-state ACM \cite{chatterjee2022polar}. The $q$-state active clock model (ACM)~\cite{chatterjee2022polar} is a natural discretization of the VM in $2d$ where particles move in $q$ equidistant angular directions with an alignment interaction inspired by the ferromagnetic equilibrium clock model. In the ACM, the hopping rate of a particle in state $\theta$ along any direction $\phi$ is $D(1-\varepsilon)$ for $\phi \in [0,2\pi]$ and $D\varepsilon$ for $\phi = \theta$. Here $\varepsilon$ is the self-propulsion ``velocity'' which indicates asymmetric diffusion and $D$ is the diffusion constant. On the contrary, in the DVM, the hopping probability of a particle with orientation angle $\theta$ along another discrete direction $\phi$ is zero as the particle always follows its orientation [see Eq.~\eqref{r}]. Hence, while the likelihood of hopping in non-preferred directions remains nonzero within the ACM (and also depends on the self-propulsion velocity of the particle), such movement is impossible within the DVM. It should also be noted that for $\varepsilon=0$, the ACM~\cite{chatterjee2022polar} transforms to a (diffusive) Brownian clock model whereas the $v_0=0$ limit makes the DVM purely passive and the $q$-state DVM reduces to the equilibrium $q$-state clock model. In light of the above discussion, it is then evident that the transverse fluctuations say in $q=4$ ACM, are stronger than the $q=4$ DVM. Transverse fluctuations in the ACM mainly originate from the nonzero hopping probability of a particle along its non-preferred directions where thermal fluctuation, through the inverse temperature $\beta$, also plays an important role as it controls the flipping dynamics. However, in the large $\beta$ limit of the ACM, similar to the small $\eta$ limit of the DVM, the probability that all particles moving in a particular direction will flip their orientation to another direction is also very small. But, unlike DVM, an ACM particle can move along a direction different than its orientation angle. This crucial difference in the hopping dynamics, as we will see, plays an essential role in the steady-state pattern formation of the DVM at low $\eta$ and $q$. 

DVM control parameters are the average particle density $\rho_0 = N/L_xL_y$, noise strength $\eta$, particle velocity $v_0=0.5$ (unless mentioned otherwise), and the measure of anisotropy $q$. According to Eq.~\eqref{discrete} a large $q$ signifies weak anisotropy while a small $q$ signifies strong anisotropy.

We performed numerical simulations of the stochastic process with parallel updates. The initial configuration is prepared homogeneously by assigning random orientations and positions to the particles as defined in  Eq.~\eqref{discrete} and Eq.~\eqref{r}, respectively. After the initialization, we let the system evolve under various control parameters for $t_{\rm eq}$ to reach the steady state followed by measurements of various quantities until the maximum simulation time $t_{\rm max}$. $t_{\rm eq}$ and $t_{\rm max}$ are functions of system size, $\eta$, and $q$. In our simulation, the maximum system size considered is $1024^2$ and we have observed that for this square domain, $t_{\rm eq}=10^5$ is sufficient for the system to reach the steady state irrespective of $\eta$ and $q$. So we take $t_{\rm eq}=10^5$ as the steady state time and perform steady state analysis up to a maximum time $t_{\rm max}=10^6$.

\section{Numerical Results}
\label{result}

\begin{figure}[!t]
    \centering
    \includegraphics[width=\columnwidth]{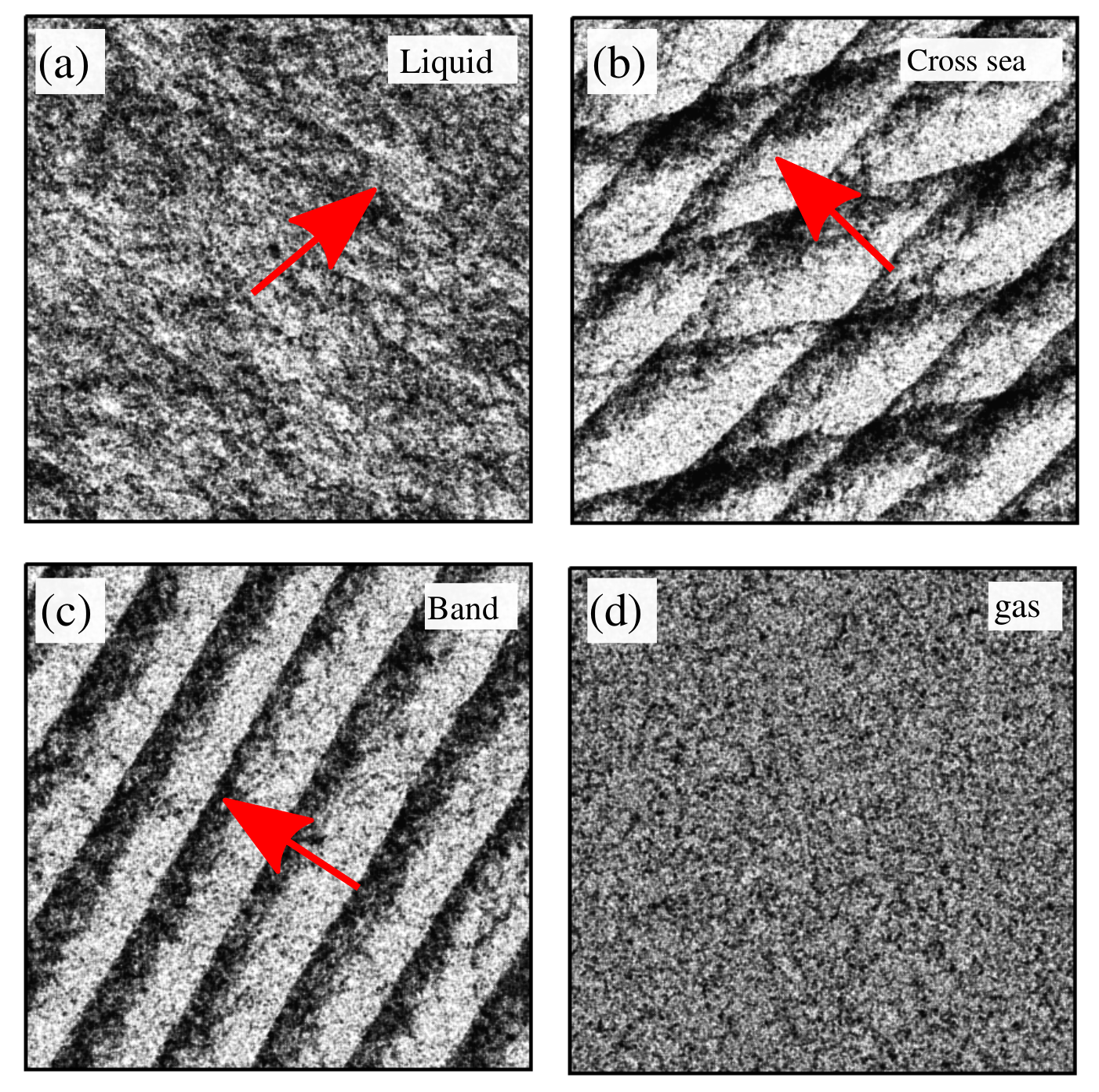}
    \caption{Snapshots on a square domain of size $1024^2$ exhibiting the four phases of the DVM for $q=9$. (a) Polar ordered liquid phase ($\eta=0.1$). (b) Cross-sea state ($\eta=0.3$). (c) Band state ($\eta=0.35$). (d) Disordered gas phase ($\eta = 0.5$). Dark color represents high particle density and red arrows indicate the average direction of motion. Parameters: $\rho_0=2$, $v_0=0.5$.}
    \label{TDVM_q9_L1024}
\end{figure}

{\bf\textit{Collective motion \& phase diagram.}} We present the typical non-equilibrium steady-state configurations of the DVM in Fig.~\ref{TDVM_q9_L1024} for $q=9$ and density $\rho_0=2$. The system exhibits phases of polar ordered liquid (a), liquid-gas coexistence (b--c), and disordered gas (d) as noise strength $\eta$ is increased from 0.1 to 0.5. The phase-coexistence region is characterized by a low-noise cross-sea phase (b) and a high-noise band phase (c). The cross-sea phase has particle density much higher at the crossing points than anywhere else and has recently been reported as the fourth phase of the VM \cite{kursten2020dry}. We notice that such patterns assemble spontaneously without an external drive at certain parameter values. Conversely, the band phase is a collection of high-density bands moving parallelly along a specific direction at a constant speed. Polar flocks [the homogeneous ordered liquid phase shown in Fig.~\ref{TDVM_q9_L1024}(a)] can be observed in a large class of active matter systems and have been considered robust to fluctuations. But recent studies have argued that liquid polar flocks are metastable to the presence of a small obstacle \cite{codina2022small} or to the nucleation of an opposite-phase droplet \cite{benvegnen2023metastability} and triggers counter-propagating dense clusters leading to the reversal of the liquid flow. In light of these observations, in the subsequent part of this paper, we will investigate the stability of the DVM liquid phase. Fig.~\ref{TDVM_q9_L1024} suggests that the system exists in distinct phases, which we will characterize next.

\begin{figure*}[!htbp]
    \centering
    \includegraphics[width=\textwidth]{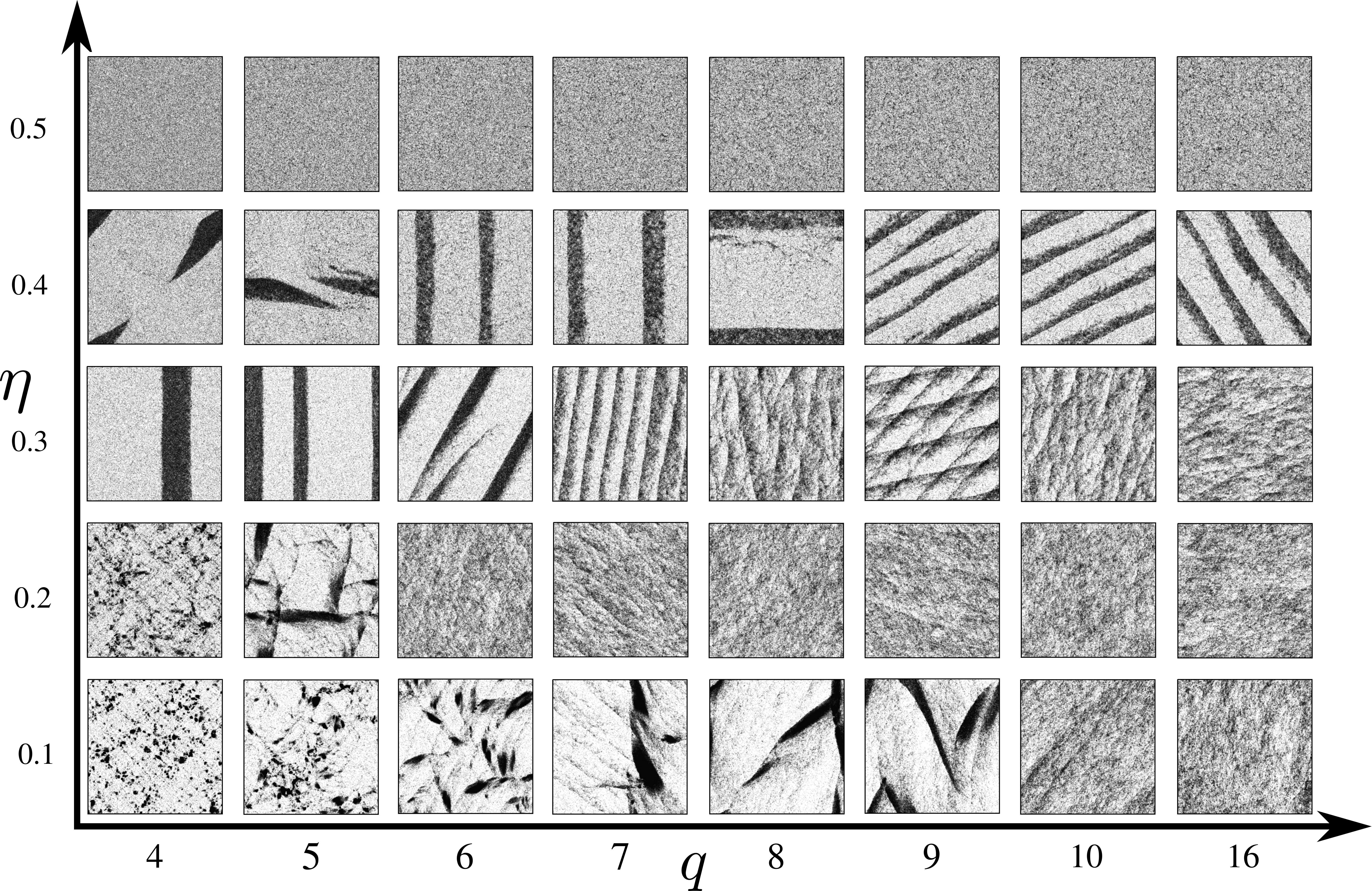}
    \caption{$\eta-q$ phase diagram of the DVM illustrated by snapshots on a $1024^2$ domain at time $t = 10^5$. As a function of $\eta$ and $q$, we observe six distinct self-organized patterns: cluster ($\eta=0.1$, $q=4$), macrophase ($\eta=0.3$, $q=4$), microphase ($\eta=0.4$, $q=6 \to 16$), cross-sea ($\eta=0.3$, $q=8 \to 10$), ordered liquid ($\eta=0.2$, $q=6 \to 16$), and disordered gas ($\eta=0.5$, $q=4 \to 16$). Parameters: $\rho_0=2$, $v_0=0.5$.}
    \label{TDVM_L1024}
\end{figure*}

The non-equilibrium steady-state behavior of the DVM is illustrated by representative late-stage snapshots as a function of noise strength $\eta$ and anisotropy parameter $q$ (see Fig.~\ref{TDVM_L1024}). We observe six distinct self-organized patterns in the ($q$, $\eta$) plane which completely describe the DVM. At low noise and $q$, we observe a locally ordered cluster phase. Although each cluster is highly dense and polar, the system as a whole does not possess any net polarization (see Appendix~\ref{appA}). This cluster phase, which has not been observed earlier in any flocking models, appears only for small $q$ and $\eta$ when the system is strongly discretized and the impact of fluctuation is insignificant. 

The appearance of a cluster phase in the $q=4$ DVM for small $(q,\eta)$ and high density instead of a polar ordered phase similar to the 4-state APM \cite{chatterjee2020flocking,mangeat2020flocking} and ACM \cite{solon2022susceptibility,chatterjee2022polar} can be attributed to the absence of transverse fluctuations through hopping. In DVM, particle movement is solely controlled by the orientation $\theta$ similar to the VM [see Eq.~\eqref{r}]. Therefore, $q=2$ DVM only manifests one-dimensional movement (along the $x$-axis) of high-density clusters of self-propelled particles having orientations $\theta = 0$ and $\theta=\pi$. We observe that these clusters never coalesce due to the lack of transverse fluctuations and thus never form a band or polar liquid phase. Similar observations are made when the constant transverse diffusion is switched off in the two-dimensional active Ising model (AIM) \cite{solon2015flocking}, although the one-dimensional AIM \cite{solon1dAIM} exhibits a flocking state consisting of a single dense ordered aggregate. Unsurprisingly, analogous to $q=2$ and $q=4$, for small $\eta$, we observe a cluster phase also for the $q=3$ DVM. We do not observe any band formation similar to $q=4$ even when noise is increased. For larger noise strengths, cluster size reduces, and the system exhibits a disordered gas phase. For fixed noise, an increase in density only increases the cluster size without changing the system morphology. This signifies that diffusion along the non-preferred hopping directions plays a crucial role in forming large liquid domains. For APM and ACM at large $\beta$ and small $q$ and DVM with small $\eta$ and $q$, the probability of transverse flipping is very small since fluctuations are weak. However, the nonzero finite hopping rates along the unbiased directions facilitate the formation of large liquid domains in the APM and ACM; whereas the absence of unbiased hopping gives rise to a cluster phase in the DVM. Thus, in principle, the interplay of $q$ and $\eta$ determines the fate of the cluster phase. If $q$ is small but fluctuation is large, the rigid cluster phase can relax and form a large ordered domain (see the snapshot for $q=6$ and $\eta=0.2$). On the contrary, if $\eta$ is small but $q$ is large, the weak anisotropy helps the cluster phase merge into a large ordered domain (see the snapshot for $q=10$ and $\eta=0.1$). This also explains why the cluster size increases with $q$ for a fixed $\eta$ (see the snapshots for $\eta=0.1$). 

Beyond the cluster phase, for $\eta=0.2$, we observe the emergence of the polar ordered liquid phase ($q \geqslant 6$). For intermediate noise strength ($\eta=0.3-0.4$), a transition is observed from macrophase separation ($q=4$, a single liquid band moving through the gaseous background) to microphase separation (multiple bands moving parallelly or in a cross-sea pattern through the gas phase) in the coexistence region where the number of bands increases with $q$. This is a consequence of having more particle orientations allowed through $q$. The cross-sea phase, where interactions become more intense due to the characteristics of the band structure, appears between the polar liquid phase and the parallel band state \cite{xue2023machine} for a fixed $q$. This phase is not simply a superposition of waves of inclined bands, but an independent self-organized complex pattern with an inherently selected crossing angle. It is worth noting that a single cross-sea pattern typically involves the crossing of at least two bands, where the crossing angle is approximately $\sim \pi/4$~\cite{kursten2020dry}. We have observed that this crossing angle remains consistent regardless of $q$. For a fixed $\eta=0.3$, as $q$ is increased from $q=7$ to $q=16$, we observe that the self-organized patterns change from bands $(q=7)$ to cross-sea pattern $(q=8$, 9, 10) followed by a polar liquid phase ($q=16$). For the same values of the control parameters, the VM $(q \to \infty)$ exhibits features similar to the $q=16$ DVM where the phase point on the $(\eta,\rho_0)$ diagram almost lies on the liquid binodal \cite{solon2015phase}. This happens because the anisotropy becomes weaker with $q$ and, for $q \geqslant 16$, the characteristic of the system becomes similar to the VM. The VM does exhibit a cross-sea phase but at a different parameter regime \cite{kursten2020dry}. For a large noise, e.g., $\eta=0.5$, we observe a disordered gas phase irrespective of $q$. 

Since the origin of different phases in the DVM depends on the spatial anisotropy, in Appendix~\ref{appB}, we show that the cluster phase is ubiquitous for small ($q,\eta$) and present a cluster size analysis of the $q=4$ DVM for varying noise. For finite system size, we found that the steady-state DVM exhibits bistability of the cross-sea phase and the band phase a within a range of the noise amplitude which results in a hysteresis (See Appendix~\ref{app_revision}). In Appendix~\ref{app_revision}, we further discussed the stability of the cross-sea phase. See also Appendix~\ref{appC} and Appendix~\ref{appD} for more discussions on the DVM phases with spatial anisotropy (rectangular domain).

Based on the above discussion, we can quantify the DVM phase diagram. We observe four different phases in Fig.~\ref{TDVM_L1024} \cite{kursten2020dry}: the ordered phase, the cross-sea phase, the band phase, and the disordered gas phase. It is, however, challenging to define the phase boundaries by visual inspection. 

\begin{figure}[!t]
    \centering
    \includegraphics[width=\columnwidth]{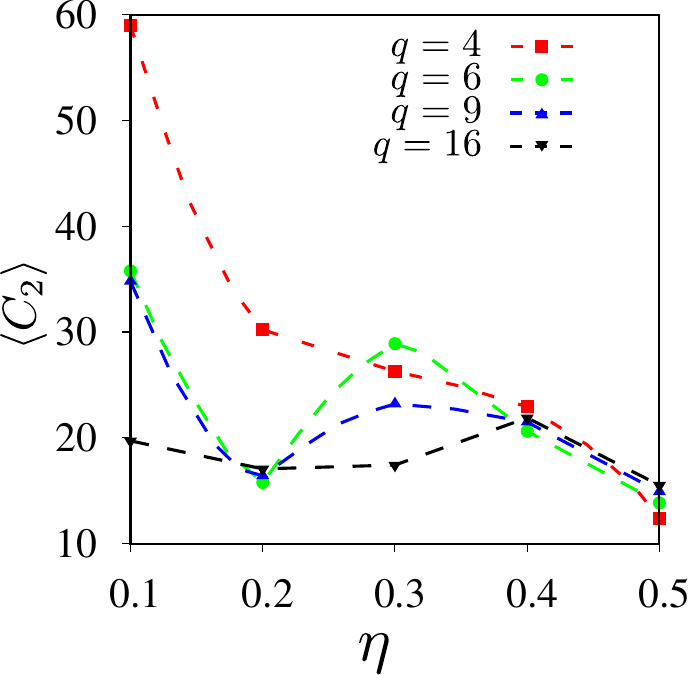}
    \caption{(Color online) The structural order parameter $\langle C_2 \rangle$ versus the noise amplitude $\eta$ for several values of $q$. The dotted curves between two successive points are approximated using spline interpolation. Parameters: $L=1024$, $v_0=0.5$, and $\rho_0=2$.}
    \label{figc2}
\end{figure} 

Previously, phase-separated density profiles were used to identify binodals that delimit the gas and liquid phases from the co-existence region~\cite{solon2015phase}. Yet, this technique can not differentiate between the distinct self-organized patterns we observe in the coexistence region (i.e. macrophase separation, microphase separation, and cross-sea) of the DVM. One might also want to distinguish different structures in terms of the global polar order parameter or the global magnetization defined as: 
\begin{align}
\label{GOP}
m = |{\bf m}| &= \frac{1}{N} \left|\sum_i {\bm \sigma}_i\right| \, .
\end{align}
The polar order parameter can be used to study the transition between the disorder gas phase and the band states but can not differentiate between the band and the cross-sea states \cite{kursten2020dry}. The Binder cumulant constructed from this order parameter shows only the transition between the disordered phase and the band state \cite{kursten2020dry}. The polar order parameter $m$ is also insensitive around the cross-sea state. Thus, for precise quantification of different phases and their boundaries, we compute the structural order parameter $C_2$ \cite{kursten2020dry,xue2023machine,kursten2020multiple} as follows:
\begin{align}
\label{c2}
C_2 &= N^2 \int G_2({\bf r}_1,{\bf r}_2)d{\bf r}_1 d{\bf r}_2 \Theta(R-\vert{\bf r}_1 \vert) \Theta(R-\vert {\bf r}_2\vert) \nonumber \\
 &= \left(\frac{N}{L^2}\right)^2 \int_{\mathbb{R}^2} \left[g(\vert {\bf r}_1-{\bf r}_2\vert)-1\right] \nonumber \\ 
& \times \Theta(R-\vert{\bf r}_1 \vert) \Theta(R-\vert {\bf r}_2 \vert) d{\bf r}_1 d{\bf r}_2 \, ,
\end{align}
where $G_2({\bf r}_1,{\bf r}_2)=P_2({\bf r}_1,{\bf r}_2)-P_1({\bf r}_1)P_2({\bf r}_2)$ for one- and two-particle probability density functions $P_1$ and $P_2$, $\Theta$ is the Heaviside step function, and $R$ is the distance around an arbitrary fixed point in space. For macroscopically isotropic systems, $G_2$ can be expressed in terms of the pair correlation function $g(r)$ which physically signifies the probability of finding a particle at a distance $r$ relative to that of a given reference particle and provides a statistical description of the local packing and particle density of the system. It has been shown \cite{kursten2020dry,kursten2020multiple} that $C_2$ performs better than $m$ in capturing the structural change and the Binder cumulant of $C_2$ is also more efficient in distinguishing different features than the Binder cumulant of the order parameter $m$. In practice, to compute $C_2$, we take all pairs of particles, draw circles of radius $R$ around them, and calculate the overlap area of the two circles. The overlap area is given by $A_{\rm overlap} = 2R^2 \cos^{-1} \left(\frac{d}{2R}\right) - \frac{d}{2} \sqrt{4R^2-d^2}$ \cite{kursten2020multiple}, where $d$ is the distance between the centers of the circles. 

\begin{figure}[!t]
    \centering
    \includegraphics[width=\columnwidth]{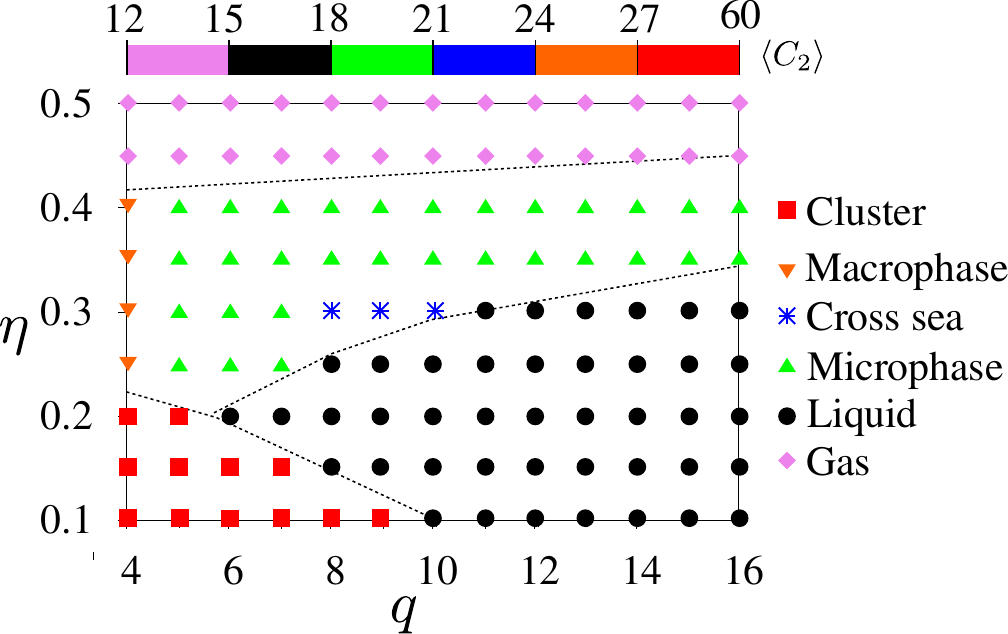}
    \caption{(Color online) $\eta-q$ phase diagram of the DVM by computing $\langle C_2 \rangle$. The colorbar represents the range of $\langle C_2 \rangle$ values for different phases. Colorbar represents the range of $\langle C_2 \rangle$ values for different phases. $\langle C_2 \rangle_{\rm gas}$: $12\to 15$, $\langle C_2 \rangle_{\rm liquid}$: $15\to 18$, $\langle C_2 \rangle_{\rm microphase}$: $18\to 21$, $\langle C_2 \rangle_{\rm cross-sea}$: $21\to 24$, $\langle C_2 \rangle_{\rm macrophase}$: $24\to 27$, $\langle C_2 \rangle_{\rm cluster}$: $27\to 60$. Parameters: $L=1024$, $v_0=0.5$, and $\rho_0=2$.}
    \label{phasediagram}
\end{figure}

The dependence of the time-averaged structural order parameter $\langle C_2 \rangle$ with respect to $\eta$ for different $q$ is shown in Fig.~\ref{figc2}. The value of $\langle C_2 \rangle$ is the lowest for disorder gas and becomes maximum when particles are clustered. For the band phase, $\langle C_2 \rangle$ values for macrophase separation are larger than the microphase separation and cross-sea phase. Between microphase separation and the cross-sea phase, $\langle C_2 \rangle_{\rm cross} > \langle C_2 \rangle_{\rm micro}$. Although a change in $\langle C_2 \rangle$ is not very significant in these two phases, it is still a better candidate for distinguishing the cross-sea from the microphase separation than the traditional polar order parameter. In Fig.~\ref{phasediagram}, we plot the $\eta-q$ phase diagram by computing $\langle C_2 \rangle$ for the six different phases where in the coexistence region, $\langle C_2 \rangle_{\rm macro}>\langle C_2 \rangle_{\rm cross} > \langle C_2 \rangle_{\rm micro}$. This phase diagram complements Fig.~\ref{TDVM_L1024}, which has been constructed using the density field and depicts the nature of phase separation in the DVM as a function of $q$. The phase diagram of Fig.~\ref{phasediagram} has been constructed on a large square domain of dimension $1024^2$. For this system size, one can expect that the finite size effect is relatively small and Fig.~\ref{phasediagram} will remain qualitatively the same in the thermodynamic limit. As discussed in Appendix~\ref{app_revision}, the DVM might show the bistability of two different coexistence states at the steady state for a particular noise amplitude. However, we believe the system will evolve to a definite steady-state phase at the thermodynamic limit $(L \to \infty)$.

Now, the nature of the coexistence region for any finite, large $q$ has been a subject of discussion in the context of $q$-state ACM \cite{solon2022susceptibility,chatterjee2022polar}. It was argued in Ref.~\cite{solon2022susceptibility} that spatial anisotropy plays a crucial role in determining the macro/microphase separation of the coexistence region in the ACM and one should observe a macrophase separation of the coexistence region for a finite $q$ beyond a characteristic length scale which diverges for large $q$. ACM with a different set of dynamical rules than Ref.~\cite{solon2022susceptibility} was studied in Ref.~\cite{chatterjee2022polar} where the flocking transition in the ACM was argued to be equivalent to the VM at large $q$. The $q$-state DVM is governed by a completely different set of microscopic rules than both the models of the ACM~\cite{solon2022susceptibility,chatterjee2022polar} and we will therefore investigate next the impact of microscopic rules on the DVM steady-state as a function of $q$. We plan to do this through the analysis of number fluctuations, the pinning-unpinning property of the system's global order, and the structure factor manifesting the correlation of polarization.

{\bf\textit{Number fluctuations.}}---It is known that fluctuations play an essential role in selecting the phase-separated patterns in flocking models. In the AIM \citep{solon2015flocking}, the density fluctuation $\Delta n^2=\langle n^2 \rangle-\langle n \rangle^2$ (where $n$ is the number of particles in a box of size $<L$) in the liquid phase was found to be normal ($\Delta n^2=n$) and the AIM coexistence region shows a macrophase separation with a single liquid domain moving on a gaseous background. In the VM \cite{solon2015phase}, on the contrary, giant density fluctuations ($\Delta n^2=n^{1.6}$) are observed which break large liquid
domains and prevent the bands from coarsening further. This results in a microphase-separated coexistence region. Giant density fluctuations of the homogeneous ordered phase have also been evaluated in other flocking models~\cite{gregoire2004onset,chate2006simple,chate2008collective,mishra2010fluctuations,ginelli2010relevance,dey2012spatial,ngo2014large}. In the $q$-state ACM \cite{solon2022susceptibility,chatterjee2022polar}, which can be thought of as a bridge between the discrete AIM and the continuous VM, density fluctuations show a transition from normal to giant fluctuation as $q$ increases and can be explained as a transition from AIM physics to VM physics. As the DVM also exhibits a transition from macrophase to microphase (and cross-sea) separation of the coexistence region as $q$ increases, it is thus useful to investigate the density fluctuation in the DVM.

\begin{figure}[!t]
    \centering
    \includegraphics[width=\columnwidth]{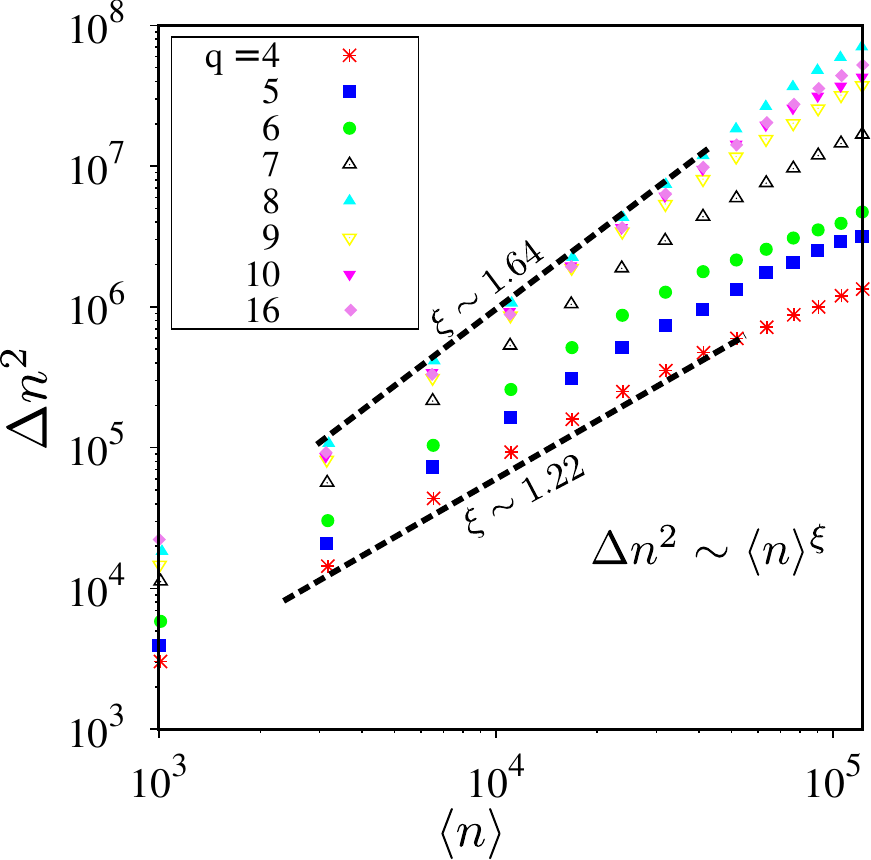}
    \caption{(Color online) Number fluctuations $\Delta n^2=\langle n^2 \rangle-\langle n \rangle^2$ versus average particle number $\langle n \rangle$ for several $q$ values in a $300^2$ domain. Parameters: $\eta=0.3$, $v_0=0.5$, and $\rho_0=6$.}
    \label{NF}
\end{figure}

We show the number fluctuations $\Delta n^2$ versus average particle number $\langle n \rangle$ in Fig.~\ref{NF}, computed in the ordered liquid phase of the DVM for various $q$. $n(\ell)$ is the number of particles in boxes of different sizes $\ell$ included in a $300^2$ domain (with $\ell \leqslant 150$), with $\langle n \rangle=\rho_0\ell^2$. As shown in Table~\ref{table_exponents}, the number fluctuation behaves like $\langle n \rangle^\xi$ with the fluctuation exponent $\xi$ increasing with $q$, from $\xi \simeq 1.22$ for $q=4$ to $\xi \simeq 1.64$ for large $q$. This transition from normal fluctuations for small $q$ to giant fluctuations for larger $q$ was also observed in the ACM~\cite{solon2022susceptibility,chatterjee2022polar} although $\xi$ for $q=4$ and 5 are moderately larger in the DVM than the ACM. The increase in density fluctuations with $q$ can be attributed to the fact that, for large $q$, particles have more rotational degrees of freedom due to the weak anisotropy and therefore more directional freedom to propel. The existence of giant number fluctuations (GNF) and its connection with microphase separation in the VM was hypothesized in Ref.~\cite{solon2015phase}. It was argued that GNF ($\xi \simeq 1.6$) breaks bulk liquid domains and produces a smectic-like microphase separation in the coexistence regime whereas the system stabilizes in the bulk phase when the density fluctuations are normal ($\xi \simeq 1$)~\cite{solon2015flocking}. Using the same logic for ACM \cite{chatterjee2022polar}, it was argued that GNF is responsible for microphase separation in the coexistence regime for $q \geqslant 8$, although a direct relation between the existence of GNF in the ordered phase and the microphase separation in the coexistence phase is still ambiguous. Nonetheless, if we compare $\xi$ in Table~\ref{table_exponents} and the snapshots in Fig.~\ref{TDVM_L1024}, we observe a correspondence between the fluctuation exponent and the pattern formation in the coexistence region of DVM. For $q<8$, although the difference in $\xi(q)$ is small, the exponents change continuously with varying $q$, similar to the active clock model~\cite{chatterjee2022polar}.

One should also consider the finite size effect on the fluctuation exponents as discussed in Ref.~\cite{chatterjee2022polar}. In Fig.~\ref{NF}, the data can be fitted to two different power-law regimes (the extracted exponents depend on the interval along the $x$-axis to which the fits are restricted): (a) $\xi$ tabulated in Table~\ref{table_exponents} in the interval $[10^3,5 \times 10^4]$ and (b) a smaller $\xi$ in the interval $[5 \times 10^4,10^5]$. Around the second interval, the plot shows a ``saturation'' because $\xi$ must decrease with increasing $\langle n \rangle$ due to the finite-size cut-off at $\langle n \rangle=N=\rho_0L^2$, where $\Delta n^2$ vanishes. 

\begin{table}

\begin{center}
\begin{tabular}{ |c|c|c|c|c|c|c|c|c| } 
\hline
$q$ & 4 & 5 & 6 & 7 & 8 & 9 & 10 & 16 \\
\hline
$\xi$ & 1.22 & 1.28 & 1.41 & 1.48 & 1.67 & 1.64 & 1.64 & 1.64 \\
\hline
\end{tabular}
\caption{(Color online) Number fluctuation exponents $\xi$ for several values of $q$, reported from Fig.~\ref{NF}. The typical error on the fluctuation exponents is 0.03. \label{table_exponents}}
\end{center}

\end{table}

{\bf\textit{Pinned property of the order parameter.}}---The global order parameter defined in Eq.~\eqref{GOP} quantifies the overall ordering of the particles in the system. In the ACM~\citep{solon2022susceptibility}, at finite size, the direction of the global polar order $\Phi \equiv \rm{arg}\langle {\bf m} \rangle$  exhibits distinct behaviors in the liquid phase depending on the value of $q$: it displays AIM-like properties (pinned along an angle) for small $q$ and VM-like behavior (unpinned over time) for large $q$. However, for a particular $q$, an unpinning to pinning transition is observed as the system size is increased which indicates the sensitivity of the ACM~\citep{solon2022susceptibility} steady-states on system size and a possible transition from micro to macro phase separation of the coexistence region beyond a length scale. We therefore are interested in analyzing this property for the DVM.

\begin{figure}[!t]
    \centering
    \includegraphics[width=\columnwidth]{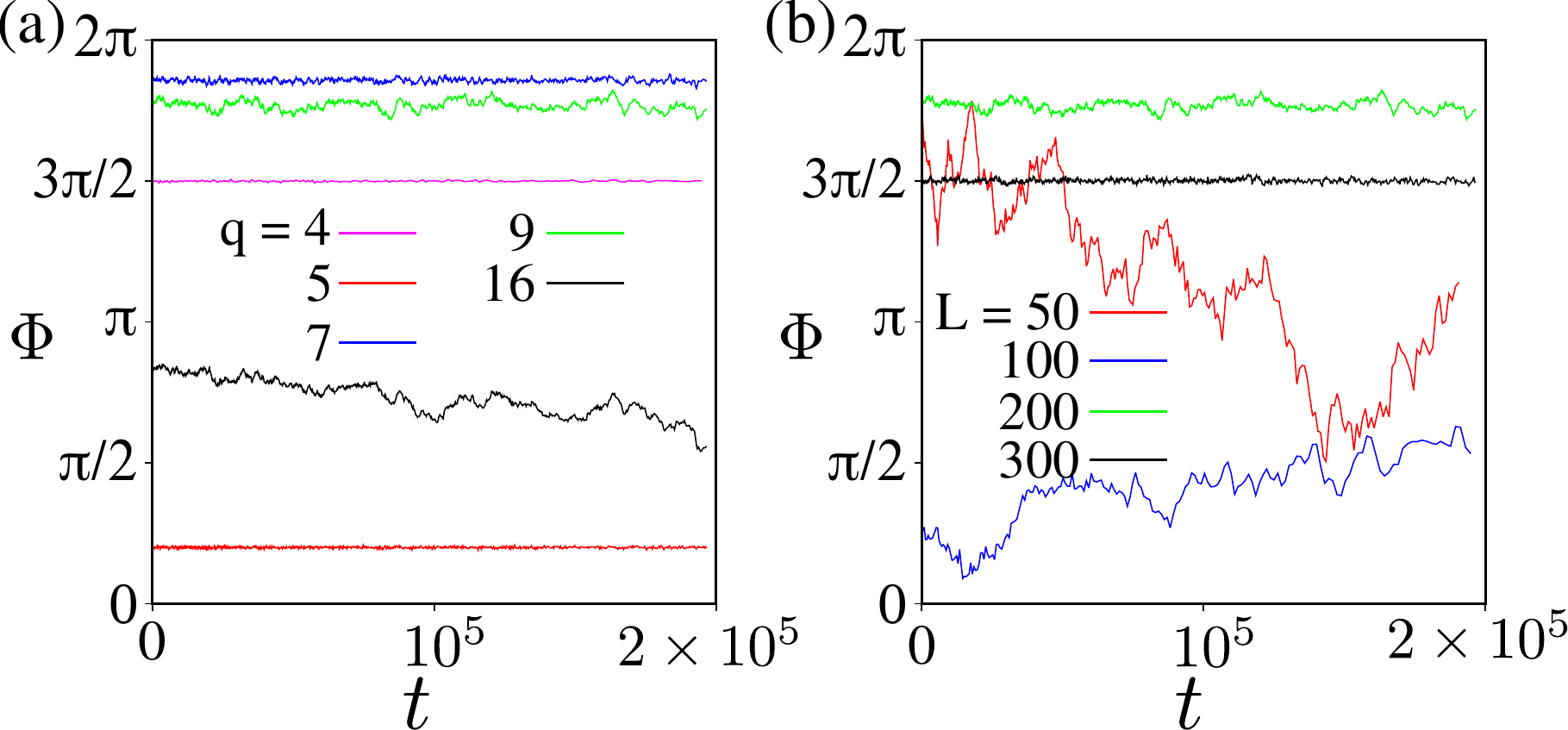}
    \caption{(Color online) (a) Time series of the orientation of global polar order $\Phi$ in the liquid phase showing a transition from unpinned to pinned as $q$ increases ($\eta=0.3$, $\rho_0=6$, $v_0=0.5$, and $L=200$). (b) $\Phi$ shows an unpinning to pinning transition as a function of system size $L$ for $q=9$ ($\eta=0.3$, $v_0=0.5$ and $\rho_0=6$).}
    \label{figGOP}
\end{figure}

In Fig.~\ref{figGOP}(a), we show the time evolution of $\Phi(t)$ in the DVM liquid phase for varying $q$. Similarly to the observation made in the ACM \cite{solon2022susceptibility}, $\Phi(t)$ begins wandering slowly with $q$ and becomes an unpinned variable of $t$ for large $q$. In other words, for weak anisotropy, the global ordering does not remain constrained to a specific orientation. While, for small $q$, $\Phi(t)$ remains pinned and tends to exhibit a stable global ordering. Microscopically, this refers to a picture in which, at large $q$, a proportional number of degrees of freedom allows the particles to choose between adjacent directions facilitated by fluctuation, while it is not the obvious choice for particles in small $q$ that require a significantly larger jump to switch directions. Translating this to the global polar order parameter and comparing Fig.~\ref{figGOP}(a) with Fig.~\ref{NF} we propose that GNF corresponds to the unpinned behavior of $\Phi(t)$. Likewise, the unpinning nature of the direction of the global polar order is a characteristic of microphase separation. 

In addition to $q$, the finite system size also affects the evolution of $\Phi(t)$ (as was also shown in Ref.\citep{solon2022susceptibility}), which is shown in Fig.~\ref{figGOP}(b) for $q=9$. Similar to the ACM \cite{solon2022susceptibility}, we observe a transition from unpinned behavior to pinned behavior in $\Phi(t)$ as the system size increases. In larger systems with polar order, a particle interacts with more particles in the neighborhood and correlates over longer distances. Higher connectivity promotes stronger alignment and cooperative motion among the particles. As a result, the direction of the global order becomes more pronounced and persistent in larger systems, leading to the pinned state. Fig.~\ref{figGOP}(b) further indicates that if $\Phi(t)$ is pinned for $L=300$, it must also be pinned for $L=1024$. This is inconsistent with the microphase separation and the cross-sea patterns observed in the coexistence region of the $q=9$ DVM (Fig.~\ref{TDVM_L1024}). Evidently, the correlation proposed earlier between the pinned property of the system's ordered liquid phase and the system morphology observed in the coexistence region (macro/micro/cross-sea) is not conclusive in the DVM (see Appendix~\ref{appF} for more details).

{\bf\textit {Structure factor.}}---To understand the system's behavior for different $q$, investigation of the magnetization correlation function which analyzes how the particle's orientations are correlated as a function of $q$ is a good measure. In the ACM \cite{solon2022susceptibility}, it was observed that the structure factor (Fourier transform of the orientation correlation function) converges to finite values at large wavelengths beyond a crossover length scale. This length scale is a function of $q$ which signifies the crucial impact of the system size on anisotropy. 

\begin{figure}[!t]
    \centering
    \includegraphics[width=\columnwidth]{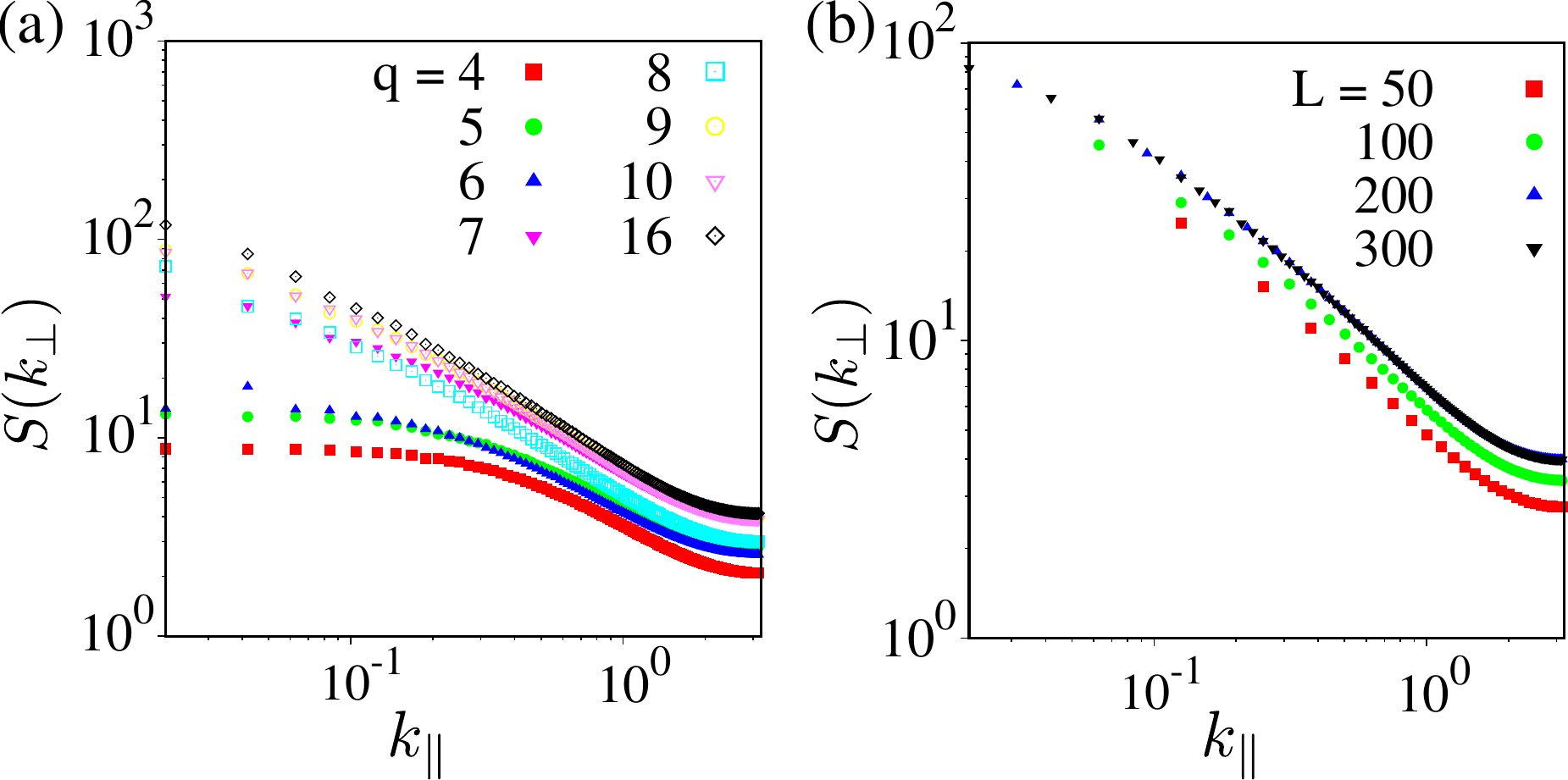}
    \caption{(Color online) Structure factor $S_{\perp}({\bf k})$ vs ${\bf k}=(k_{\parallel},0)$ in the ordered liquid phase for (a) several $q$ values ($L=300$) and (b) varying system size ($q=9$). Parameters: $\eta=0.3$, $\rho_0=6$, and $v_0=0.5$.}
    \label{SF}
\end{figure}

To explore the correlation among particle polarizations, we consider the transverse magnetization structure factor $S_{\perp}({\bf k})=\langle m_{\perp}({\bf k}) m_{\perp}(\bf{-k}) \rangle$ \cite{solon2022susceptibility} against wavelength ${\bf k}$ and plot it in Fig.~\ref{SF}. The structure factor has been calculated in the liquid phase on a $300^2$ domain for (a) various $q$ values (the same behavior is observed for the structure factor of the density field) and (b) for a fixed $q=9$ but for various system sizes. The results presented in Fig.~\ref{SF}(a) show that for small $q$, the structure factor $S(k)$ converges to finite values as the wave vector ${\bf k} \to 0$. This convergence indicates an AIM-like behavior or a macrophase separation of the coexistence region \cite{solon2022susceptibility}. However, one can notice that this convergence is achieved only beyond certain length scales and these length scales are functions of $q$. The structure factor captures the correlations between particle orientations and therefore the magnitude of ordering at different length scales. A saturation of $S(k)$ for small $q$ when ${\bf k}$ approaches zero thus signifies a strongly correlated liquid domain whereas for large $q$, due to weak anisotropy or more allowed orientations for ordering, particles inside the liquid domain are not as strongly correlated as for small $q$. Fig.~\ref{SF}(b) shows $S(k)$ for several system sizes and manifests that with larger system sizes $(L \geq 300)$, $S(k)$ tends to converge to a finite value when ${\bf k} \to 0$. Our earlier argument that stronger interactions between particles (with larger $L$) promote robust ordering also applies here. Comparing Fig.~\ref{SF} with Fig.~\ref{figGOP}, we conclude that the pinning behavior (unpinning behavior) of $\Phi(t)$ and the saturation of $S(k)$ (algebraic scaling of $S(k)$) compliment each other and convey the same physics. 

It was argued in Ref.~\citep{solon2022susceptibility} considering the pinning properties of the order parameter and behavior of the structure factor in the liquid phase that for large $q$ ACM, VM behavior (microphase separation) will be observed only up to large finite sizes. But the asymptotic large length scale behavior will be AIM-like (macrophase separation) where the length scale diverges with $q$ as $\exp(q^2)$. Ref.~\citep{solon2022susceptibility} also showed that in the phase coexistence region of the ACM, a microphase to macrophase transition occurs when the linear system size increases along the transverse direction at fixed $q$, which we do not observe (in the DVM, multiple bands do not merge to a single band when $L_y$ is increased for a fixed $L_x$). We observe a similar behavior of $\Phi(t)$ and $S(k)$ as Ref.~\citep{solon2022susceptibility} but can not draw a conclusive correspondence between the large length-scale liquid phase behavior of $\Phi(t)$ and $S(k)$ to the phase-coexistence behavior of the DVM as shown in Fig.~\ref{TDVM_L1024}.

In Fig.~\ref{TDVM_L1024}, the snapshots on a large square domain (without any spatial anisotropy) show the existence of a microphase separation and cross-sea patterns (for which one needs at least two bands) of the phase-coexistence region for large $q$ values. The number fluctuation plotted in Fig.~\ref{NF} corroborates this observation by exhibiting GNF for those large $q$ values. The large length scale asymptotic behavior (for large $q$) of the direction of global order $\Phi(t)$ (Fig.~\ref{figGOP}) and the structure factor $S(k)$ (Fig.~\ref{SF}) in the ordered liquid phase respectively shows a pinned behavior and saturation for $k_{\parallel}=0$ which as argued in Ref.~\citep{solon2022susceptibility} signifies an AIM phenomenology. However, our numerical investigation of the DVM does not show a cross-over from micro- to macrophase separation for higher $q$ values as observed in the ACM~\citep{solon2022susceptibility} although matches with the observation of another model of ACM~\citep{chatterjee2022polar} with different dynamical rules. Therefore, we argue that the impact of dynamical rules governing flipping and hopping in a flocking model has a significant influence over the system dynamics.

\begin{figure}[!t]
    \centering
    \includegraphics[width=\columnwidth]{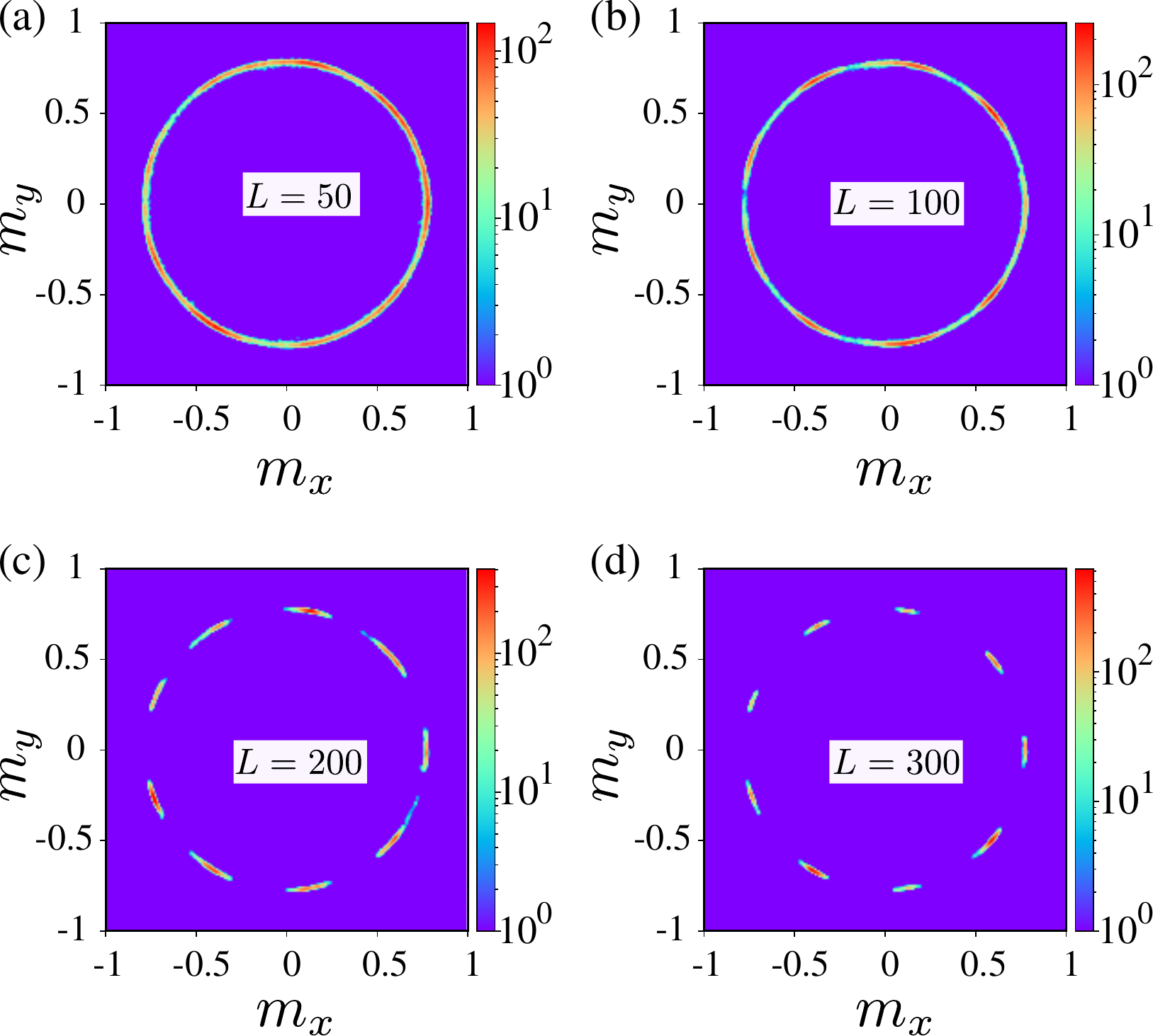}
    \caption{(Color online) Order parameter distributions of the $q=9$ DVM for the liquid phase. Parameters: $\eta=0.3$, $v_0=0.5$ and $\rho_0=6$. Ring-like distributions for smaller system sizes ($L=50$ and $L=100$) in (a) and (b) change to distinct isolated spots for larger system sizes ($L=200$ and $L=300$) in (c) and (d), which correspond to the ordered liquid phase's $q$-fold degeneracy.}
    \label{op_distro}
\end{figure}

{\bf\textit{Order parameter distribution.}}---In Vicsek-like models, where particles are active and move with a constant velocity, the ordered state exhibits a true long-range order (LRO) in two dimensions because of a spontaneous symmetry breaking due to the out-of-equilibrium nature of the models. In Fig.~\ref{op_distro}, we show the time- and ensemble-averaged distribution of the order parameter $\textbf{m}=(m_x,m_y)$ for increasing system sizes, where $m_x=\frac{1}{N}\sum_{i=1}^N \cos \theta_i$ and $m_y=\frac{1}{N}\sum_{i=1}^N \sin \theta_i$. In Fig.~\ref{op_distro}(a) and Fig.~\ref{op_distro}(b), ring-like distributions (unpinned orientations) are the characteristic of the quasi long–range ordered (QLRO) phase. However, this occurs due to the finite-size effect and is similar to the impact of finite-size on $\Phi(t)$ and $S(k)$. We recover the LRO for larger system sizes ($L=200$ and $L=300$) where the distributions display nine distinct isolated spots (pinned orientations) that correspond to the 9-fold degeneracy of the ordered liquid phase, each spot having equal probability. One can expect that the finite size effect will be much weaker for $L=1024$ and the spread of the distribution in the LRO phase around the allowed ordering angles will also be more precise.

The DVM for large $q$ exhibits a QLRO phase when $v_0=0$ (see Appendix~\ref{appE}). We argue that DVM with immobile particles reduces to the two-dimensional $q$-state clock model (with a quenched bond-disorder as only particles within a fixed distance interact) which approaches the XY model for large $q$ with vanishing LRO regime~\cite{clockmodel2018}. For $v_0>0$ and a fixed $L$, as flocking directions increase with $q$, we again observe ring-like distributions for large $q$ (see Appendix~\ref{appE}) but beyond a length scale which is proportional to $q$, the order parameter distributions for large $q$ show $q$ isolated spots characteristic of the LRO phase. This is similar to the unpinned to the pinned transition of $\Phi(t)$ and convergence of $S(k)$ to a finite value at ${\bf k} \to 0$ for large $q$ values as the system size is increased. In the VM ($q \to \infty$), even for a large $L$, the order parameter distribution shows a ring-like structure because of the continuous symmetry. 

{\bf\textit{Stability of the ordered liquid phase.}}---
\begin{figure*}[!htbp]
    \centering
    \includegraphics[width=\textwidth]{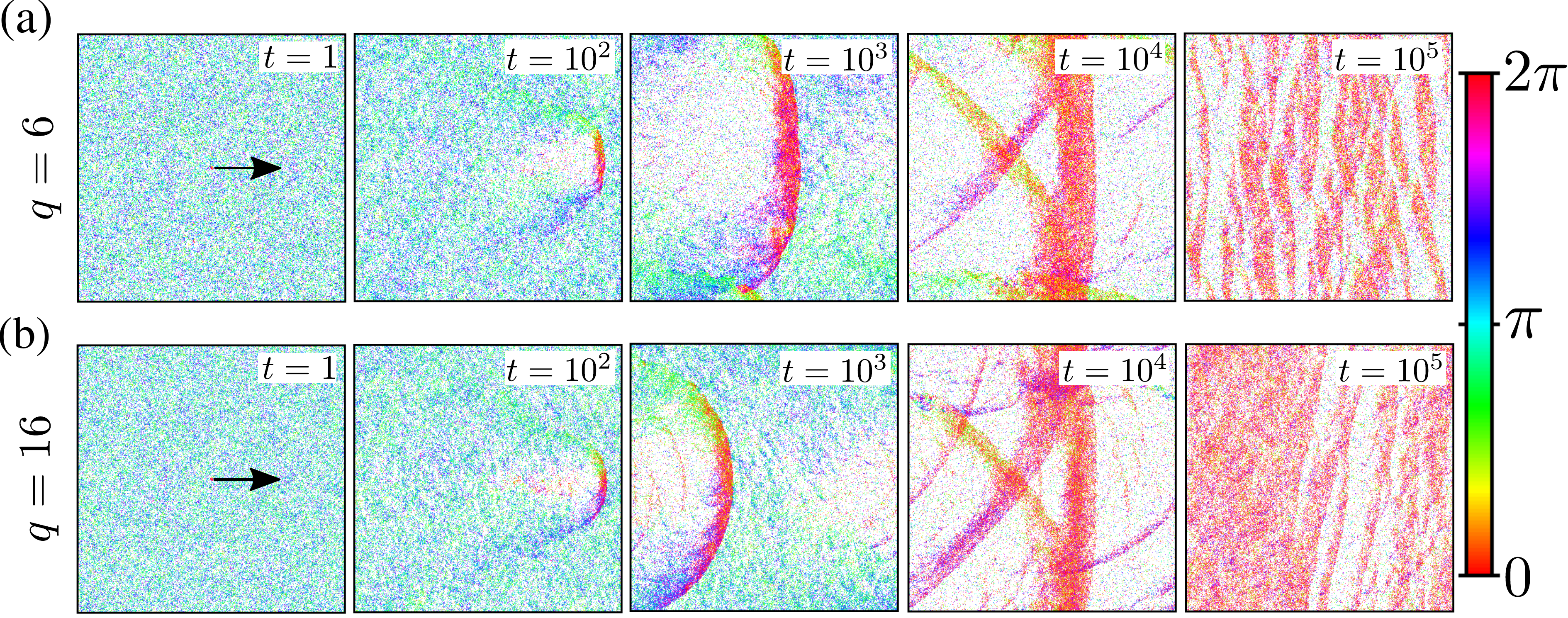}
    \caption{(Color online) Time evolution snapshots of the orientation field showing the reversal of the initial ordered liquid phase in (a) $q=6$ and (b) $q=16$-state DVM following the introduction of a tiny counter-propagating high-density droplet (motion of direction is denoted by black arrows, $t=1$) of radius $r_d=5$ and density $\rho_0^d=10\rho_0$. Colorbar represents particle orientation field. Parameters: $L=1024$, $\eta=0.4$, $v_0=1$ and $\rho_0=2$.}
    \label{metastable}
\end{figure*}
As discussed in the context of Fig.~\ref{TDVM_q9_L1024}, here we present a brief analysis of the stability of the DVM ordered liquid phase by inserting a small high-density counter-propagating liquid droplet. Initially, the average direction of the particles in the polar liquid phase is aligned along the direction $\Phi_{\rm liq}=\pi$, and the average particle orientation of the liquid droplet is $\Phi_d=0$. The radius of the droplet is $r_d$ and density is $\rho_0^d$ and it is inserted in an ordered phase of density $\rho_0$ ($\rho_0^d\gg\rho_0$). We take several $q$ values and calculate the probability ($P_{\rm rev}$) that the droplet grows against the main order and reverses the ordered phase as a function of $\eta$, $r_d$, and $\rho_0^d$. The simulation protocol follows Ref. \cite{benvegnen2023metastability}, where the initial configuration is prepared by particles with $\theta=\pi$ and we let the system evolve up to a certain time $t$ to reach the steady state. The system retains the average global polarization in the direction $\theta=\pi$. Then, a circular region of radius $r_d$ centered at $(L/2, L/2)$ is selected and an additional $\Delta N = (\rho_0^d-\rho_0)\pi r_d^2$ number of particles are added to make a high-density circular droplet. Finally, the orientation of all the particles within the droplet is changed to $\theta=0$.  

In Fig.~\ref{metastable}, we study the fate of polar flocks in $q=6$ and 16-state DVM by introducing a small high-density counter-propagating droplet against the initial polar ordered liquid phase of the main flow and observe the subsequent time evolution. One should note here that the perturbation through the droplet is very small i.e. the ratio of droplet diameter to the linear length of the simulation box is $\sim 10^{-2}$ ($r_d=5$, $L=1024$). We observe, similar to Ref.~\cite{codina2022small}, that the droplet grows with time leaving behind a dilute region ($t=10^2$) and adds more and more particles as it moves along forming a principal dense, curved band (followed by several other curved bands) that invades the whole system ballistically ($t=10^3$ and $t=10^4$). In the final stage, this principal curved band connects itself over the system boundaries and widens until a steady state liquid phase of a different $\Phi_{\rm liq}$ emerges (at $t=10^5$), signifying the metastability of the DVM liquid phase. The time-evolution is similar for both small and large $q$, which signifies that both discrete and continuous-symmetry flocks are metastable.

The growth pattern of the DVM droplet is similar to the VM~\cite{codina2022small} but distinct from the AIM \cite{benvegnen2023metastability}. In the DVM, after its introduction, the droplet front interacts with the liquid particles outside and creates a curved band of particles having several different orientations (impact of $q$). In this process, the droplet seizes to exist and it is this high-density curved band that destroys the initial flow. In AIM \cite{benvegnen2023metastability}, the droplet grows along the direction transverse to the propagation (due to the constant transverse diffusion) creating a comet-like trail of disordered particles that can not be observed in the DVM.

\begin{figure}[!t]
    \centering
    \includegraphics[width=\columnwidth]{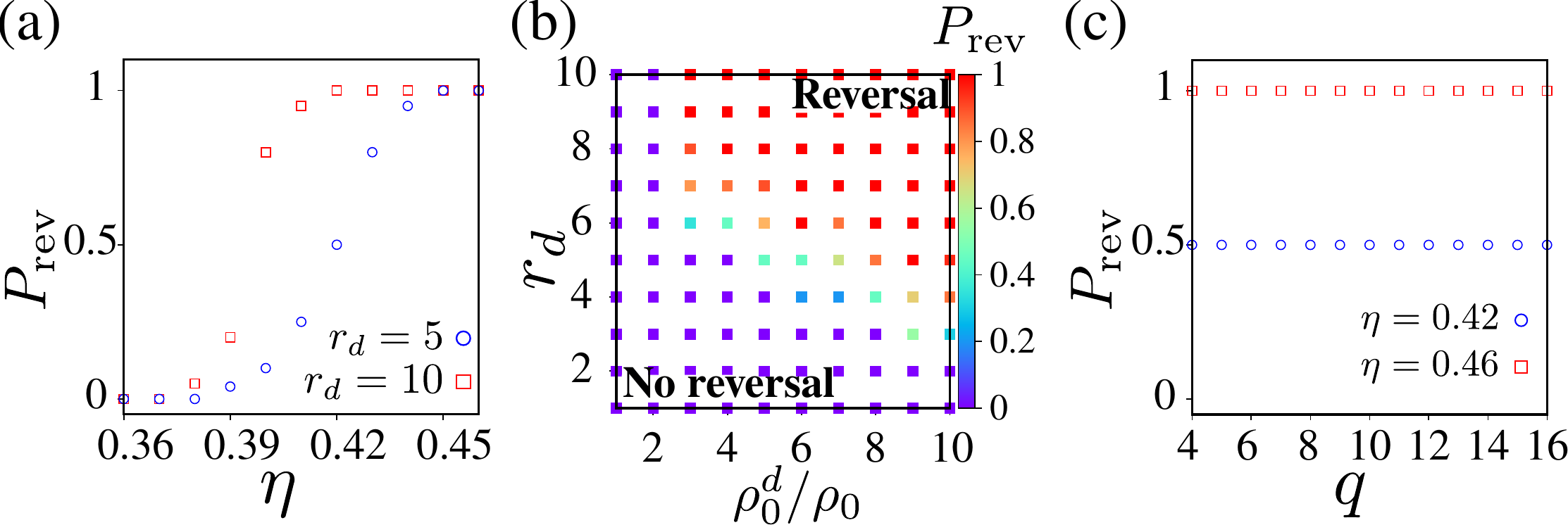}
    \caption{(Color online) (a) Reversal probability $P_{\rm rev}$ versus noise $\eta$ for $q=16$ and two different droplet radius $r_d=5$ and 10. (b) $r_d-\rho_0^d/\rho_0$ phase diagram for $\eta=0.45$ and $q=16$. The colorbar represents reversal probability $P_{\rm rev}$. (c) $P_{\rm rev}$ versus $q$ for various noise, $r_d=5$. Parameters: $L=400$, $\rho_0^d=10\rho_0$ (a \& c), $v_0=1$ and $\rho_0=2$.}
    \label{metastable2}
\end{figure}

Fig.~\ref{metastable2} quantifies $P_{\rm rev}$, the probability of reversing the main flow upon the introduction of a given droplet, for several control parameters. For each set of control parameters, we have taken 20 independent realizations to calculate $P_{\rm rev}$. Akin to Ref.~\cite{codina2022small}, we observe that the noise strength $\eta$ has a strong influence on the reversal dynamics and $P_{\rm rev}$ increases from 0 to 1 as $\eta$ is increased [Fig.~\ref{metastable2}(a)]. This is because for small $\eta$, the ordered phase is very stable, and thus, the counter-propagating dense bands find it difficult to reverse its flow. For large $\eta$, fluctuations are stronger, and therefore, the
probability of reversal increases. Fig.~\ref{metastable2}(a) also exhibits that the transitional value of $\eta$ ($P_{\rm rev}=\frac{1}{2}$) above which a droplet triggers a reversal is a decreasing function of $r_d$ although there is a critical $\eta$ ($\eta \sim 0.35$), below which no droplet can trigger a reversal irrespective of its density. The $r_d-\rho_0^d$ phase diagram in Fig.~\ref{metastable2}(b) has been constructed by calculating $P_{\rm rev}$ for several $(r_d,\rho_0^d)$. Unsurprisingly, large $r_d$ combined with large $\rho_0^d$ are shown to facilitate the reversal of the initial liquid phase. The droplet-induced reversal of the liquid flow is found independent of $q$ where $P_{\rm rev}$ is found to behave similarly for each $q$ under certain values of $\eta$ [Fig.~\ref{metastable2}(c)]. 

In addition, a study of the metastability of the DVM liquid phase on rectangular geometry ($L_x=800, L_y=100$), produces a similar outcome. One can also consider droplet movement in other directions than opposite to the global flow, e.g. transversely propagating droplets with $\Phi_d=\pi/2$ for $q=8$ or $\Phi_d=2\pi/3$ for $q=6$ and perform a similar analysis to check whether the liquid phase is susceptible to droplets irrespective of their propagation direction.

\section{Discussion}
\label{discussion}
Our study motivated by the active clock model \citep{chatterjee2022polar,solon2022susceptibility}, considers a true $q$-state discrete version of the Vicsek model where $q$ defines the strength of orientation anisotropy. At small $q$, the system is highly anisotropic, which, however, vanishes in the limit $q\rightarrow \infty$ when we recover the Vicsek model. The DVM shows qualitatively similar features as the ACM \cite{chatterjee2022polar} for intermediate noise strength $\eta$ where a transition from macrophase to microphase separation is observed in the coexistence region as $q$ increases. But for small $q$ and $\eta$, the liquid phase appearing in the ACM at low temperatures is replaced in the DVM by a cluster phase. The cluster phase consists of multiple clusters with different polarization (see Fig.~\ref{appfig1}) which does not exhibit a long-range order. The clusters grow and merge with increasing $q$ leading to a homogeneous ordered phase at large $q$. For small $q$, a long-range ordered phase can be achieved by increasing the noise strength. At low noise and small $q$, the flipping probability is very small, and in addition, transverse fluctuations through hopping are also absent. The combined influence of these factors results in clusters failing to grow continuously at small $\eta$ and $q$, preventing the system from reaching a homogeneous liquid state. Therefore, the polar ordered phase which is ubiquitous in discretized flocking models (such as the AIM, APM, and the small $q$ limit of the ACM) for small noise has been replaced by a cluster phase in the DVM and is a consequence of the strong anisotropy through $q$. The DVM recovers the ordered phase for small noise at the large $q$ limit, which signifies the DVM to be in the same class as the VM ($q \to \infty$ DVM).

The self-organized patterns in the coexistence region of the discretized VM indicate a transition from AIM-like patterns to VM-like patterns as anisotropy becomes weaker. This observation is corroborated by the giant density fluctuations for large $q$. However, the large length scale behavior of the direction of global order $\Phi(t)$, the structure factor $S(k)$, and the order parameter distribution in the liquid phase do not correspond with the phase-coexistence patterns of the large $q$ DVM. The DVM without any spatial anisotropy and at large length scales shows a transition from a macrophase-separated coexistence region to a coexistence region having microphase or cross-sea pattern with increasing $q$. This is similar to the observation made in the $q$-state active clock model \cite{chatterjee2022polar} which also approaches the VM at the large $q$ limit.

We also find that the DVM liquid phase is susceptible to perturbation applied through a counter-propagating droplet.  The liquid phase reorients and propagates along the direction of the droplet. The reversal dynamics is significantly impacted by the noise strength $\eta$ ~\cite{codina2022small} but remains independent of $q$. The stability of the high-density flocking ordered phase at low noise is still an open problem and will be addressed in a subsequent study \cite{swarnajit2023metastability}.

As a final remark, we add that the rotational flexibility of the particles and microscopic details of the dynamical rules can significantly impact the macroscopic properties of the ordered phase. It would be interesting to compare the model predictions of the DVM with suitable experiments where anisotropy in the particle orientation may be controlled by a finite number of motility directions. 

\section{Acknowledgments}
MK acknowledges financial support in the form of a research fellowship from CSIR, Govt. of India (Award Number: 09/080(1106)/2019-EMR-I). RP, MK thanks the Indian Association for the Cultivation of Science (IACS) for the computational facility. SC and HR are financially supported by the German Research Foundation (DFG) within the Collaborative Research Center SFB 1027. SC and MK wants to thank Dr. Matthieu Mangeat for many valuable discussions.

%\clearpage

%\newpage
%\onecolumngrid
\appendix
\section{Dependency of the cluster phase on the initial condition}
\label{appA}
\begin{figure}
    \centering
    \includegraphics[width=\columnwidth]{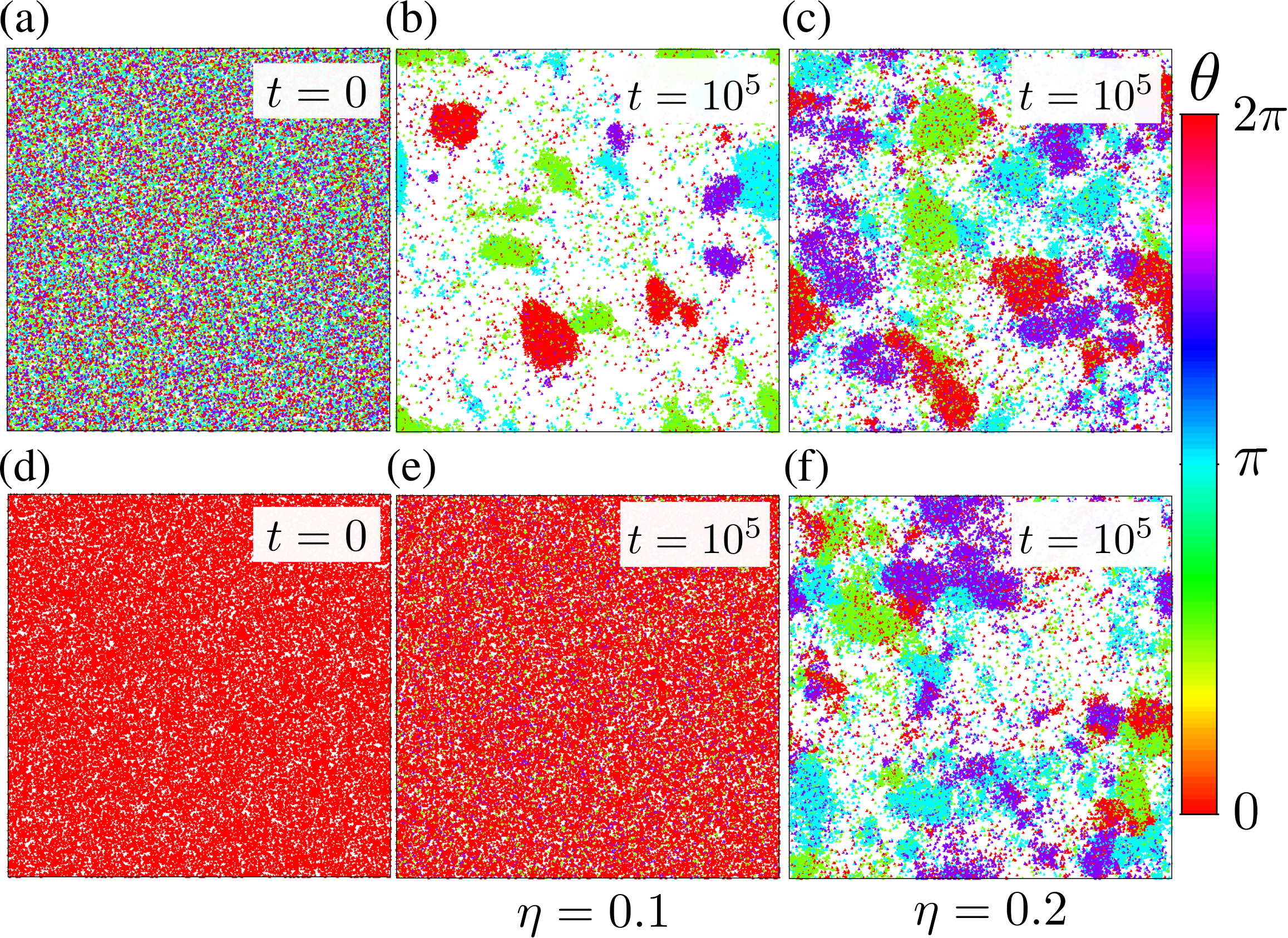}
    \caption{(Color online) Steady-state snapshots (b-c) and (e-f) from two different initial conditions: (a) random disordered and (d) polar ordered. (b, e) $\eta=0.1$ and (c, f) $\eta=0.2$. The colorbar represents particle orientations. Parameters: $q=4$, $L=300$, $v_0=0.5$, and $\rho_0=6$.}
    \label{appfig1}
\end{figure}
Snapshots in Fig.~\ref{appfig1}(b-c) and Fig.~\ref{appfig1}(e-f) illustrate the steady state behavior of the $q = 4$ DVM starting from two different initial conditions: (a) random disordered and (d) polar ordered. In the top panel, the system exhibits cluster phase for $q=4$ for both values of $\eta$ when starting from an unbiased random configuration. The initial coarsening process forms clusters, but they do not merge to create a single large ordered domain due to the absence of transverse fluctuations as discussed in the context of Fig.~\ref{TDVM_L1024}. In the bottom panel, this observation changes for the lowest noise strength when starting from a polar-ordered initial configuration. For $\eta=0.1$, the steady state remains in an ordered liquid phase (similar to the steady state behavior of the 4-state APM or ACM at low temperature) signifying that the fluctuation is weak to alter the broken symmetry phase into a cluster phase. However, with an increase in the noise ($\eta=0.2$), the steady-state cluster phase appears again. This also suggests that for DVM with weak anisotropy and fluctuations, the cluster phase is the non-equilibrium steady-state and the well-known small $\eta$ or large $\beta$ ordered liquid phase can only be achieved by taking a strongly polarized ordered initial condition. It is worth noting that the number of clusters in (c) is higher than in (b) due to more relaxation via the noise. A further increase of the noise will lead to a single large ordered domain as shown in the phase diagram of Fig.~\ref{appfig3}.

\section{Cluster size analysis for $q=4$}
\label{appB}
\begin{figure}
    \centering
    \includegraphics[width=\columnwidth]{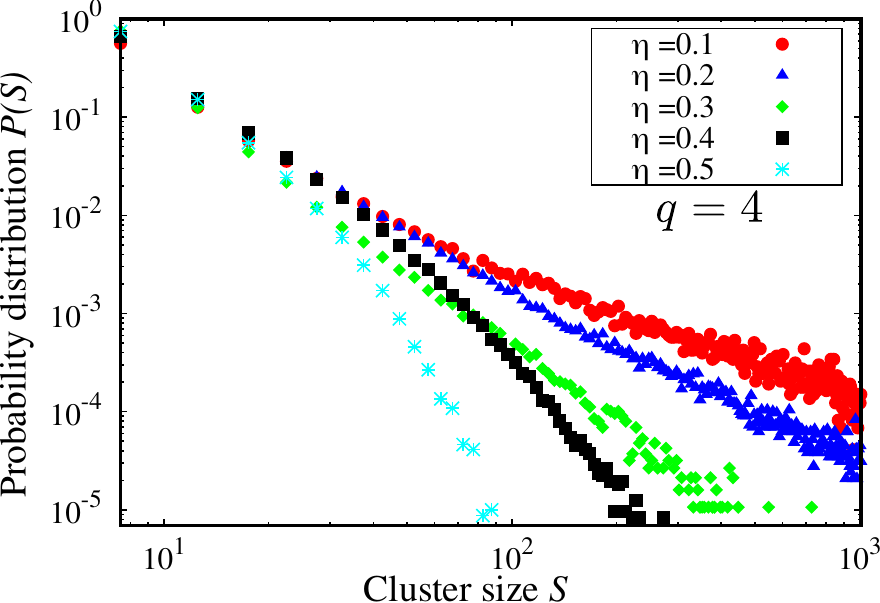}
    \caption{(Color online) Probability distribution $P(S)$ of cluster size $S$ for different noise level. Parameters: $q=4$, $L_x=800, L_y=100, \rho_0=2$, and $v_0=0.5$.}
    \label{appfig4}
\end{figure}
The cluster size analysis for $q=4$ is shown in Fig.~\ref{appfig4} for a rectangular domain of size $800 \times 100$. We use a box-counting method to find a cluster which is described below:

If $L_x$ and $L_y$ are the linear sizes of a rectangular domain, $L_x \times L_y$ is the total number of boxes we consider. For a box $i$, $n_i$ is the number of particles in that box and $c_i$ is the cluster label of the box. If the box does not contain any particles $c_i = 0$ i.e. it is not part of any cluster. Then we use the Depth-first search (DFS) algorithm to find the connected boxes that are not void of particles and label them as a single cluster. For each cluster label $c_j$, we calculate the size of the cluster $S_{c_i}$ as following:
\begin{equation}
    S_{c_i} = \sum_{j=1}^{L_x L_y} \delta_{c_i, c_j} \times n_j \, ,
\end{equation}
where $\delta_{c_i, c_j}$ is the Kronecker delta function that equals 1 if $c_i = c_j$, and 0 otherwise. Then we calculate the cluster size probability distribution denoted as $P(S)$:
\begin{equation}
    P(S) = \frac{\text{Number of clusters with size } S}{\text{Total number of clusters}} \, .
\end{equation}

Fig.~\ref{appfig4} illustrates that for small noise, the probability of larger cluster formation is high. This is because reduced fluctuation in the system facilitates the formation of high-density clusters. However, with noise ($\eta=0.3$), the cluster phase vanishes and the $q=4$ DVM exhibits a macrophase separation, resulting in a decrease in the probability of obtaining large clusters. At large noise ($\eta \geqslant 0.4$), the system becomes disordered, leading to a lesser probability of formation of large clusters. We would also like to mention that for a fixed noise, $P(S)$ versus $S$ for various orientations ($\theta=0, \pi/2, \pi, 3\pi/2$) shows almost identical distributions signifying no preference in orientation in the cluster formation.

\section{Stability of the DVM steady-state phases}
\label{app_revision}
\begin{figure*}
    \centering
    \includegraphics[width=0.8\textwidth]{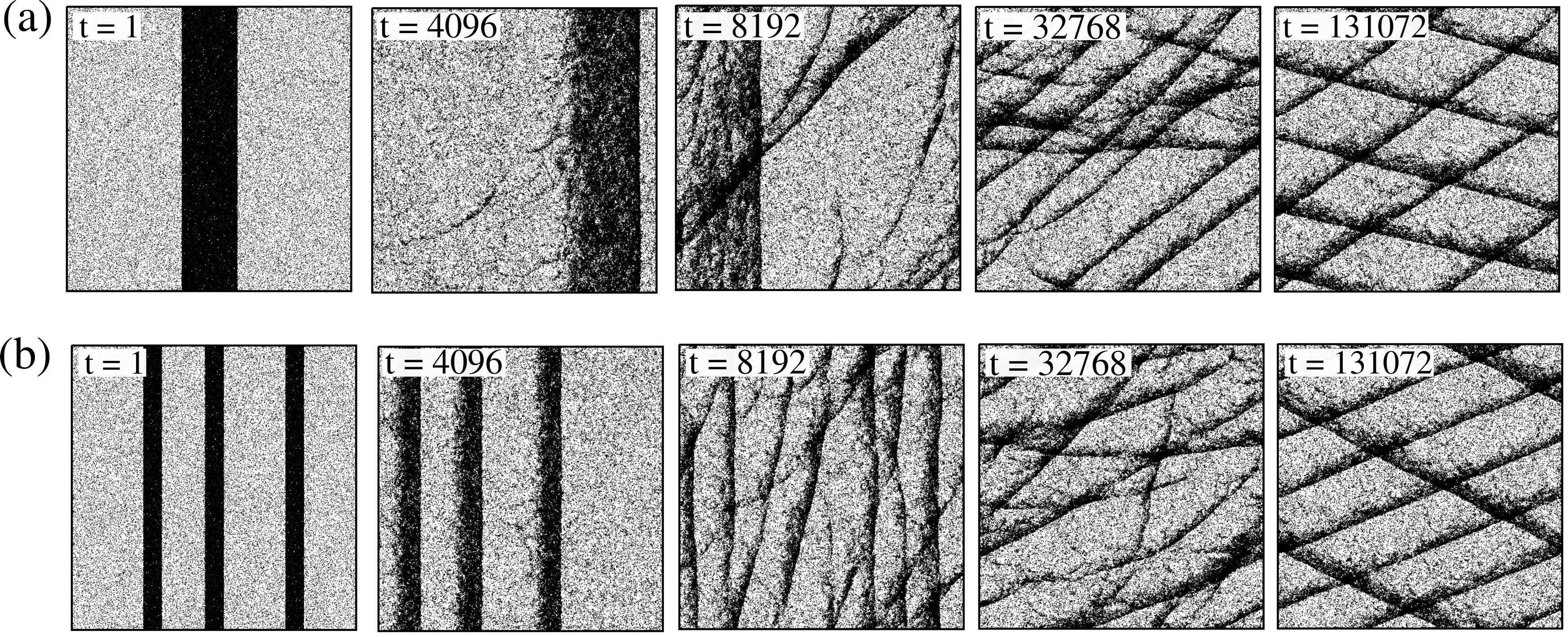}
    \caption{Stability of the cross-sea phase. (a-b) Starting from a high-density macrophase and microphase-separated coexistence band, the system finally exhibits the cross-sea phase at a large time respectively. Parameters: $L=1024$, $q=9$, $\eta=0.3$, $\rho_0=0.63$, $v_0=1$.}
    \label{stabilityfig1}
\end{figure*}
\begin{figure*}
    \centering
    \includegraphics[width=\textwidth]{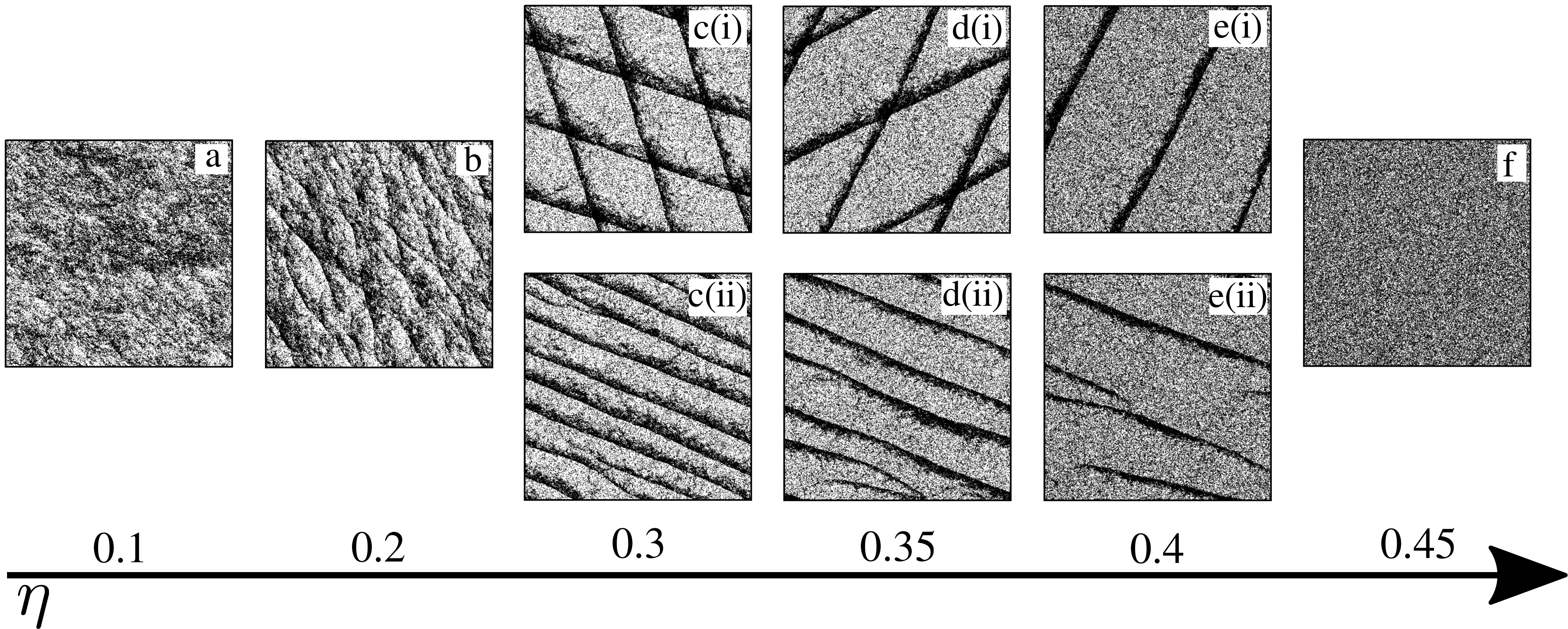}
    \caption{Steady-state snapshots of the DVM ($q=9$) at different noise amplitudes. (a-b) Ordered phase at $\eta=0.1$ and 0.2; (c-d) Bistable phase at $\eta=0.3$ and $\eta=0.35$ [(i) cross-sea phase, and (ii) microphase]; (e) microphase at $\eta=0.4$ [(i) fully developed microphase, and (ii) partial microphase]; (f) disordered phase at $\eta=0.45$. System size $L=1024$. Parameters: $\rho_0=0.63$, $v_0=1$.}
    \label{stabilityfig2}
\end{figure*}
\begin{figure}
    \centering
    \includegraphics[width=0.8\columnwidth]{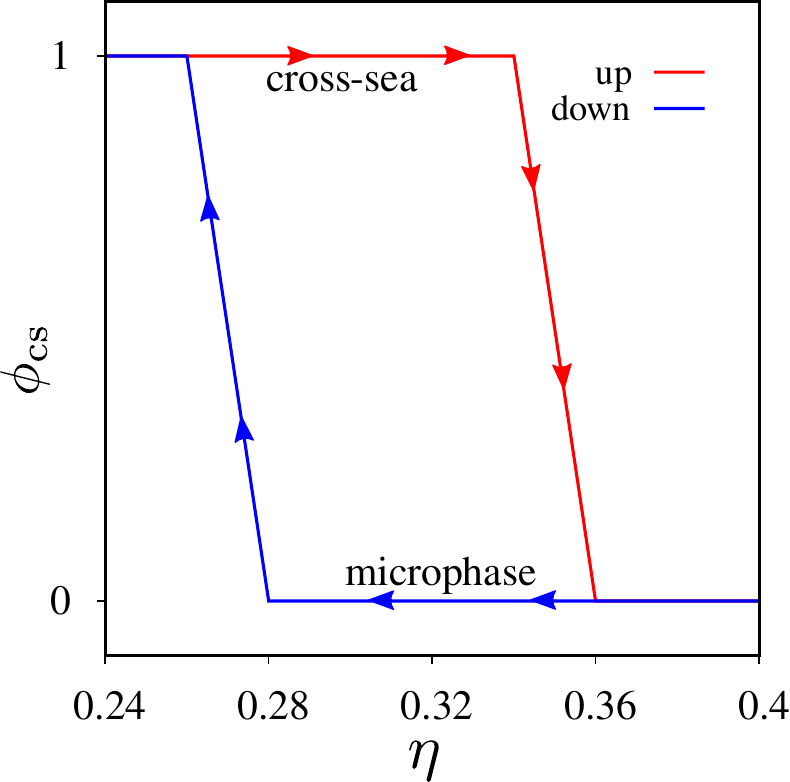}
    \caption{(Color online) Hysteresis loop in the $q=9$ DVM between the cross-sea phase and microphase-separated band phase where $\phi_{\rm cs}$ denotes the presence of a cross-sea phase ($\phi_{\rm cs}=1$). System size $L=1024$. Parameters: $\rho_0=0.63$, $v_0=1$.}
    \label{stabilityfig3}
\end{figure}

In this section, we analyze the stability of the different DVM steady-state phases. Figure~\ref{stabilityfig1} (a) illustrates the stability of the cross-sea phase where we ask if one
starts with a band state in the cross-sea regime, will the system evolve to a cross-sea pattern? The figure shows that the system initially configured in a high-density single band (visualized as a vertical stripe of aligned particles), evolves into a cross-sea phase after a long time ($t=131072$). A similar observation is made with microphase-separated initial bands in Fig.~\ref{stabilityfig1} (b), suggesting that the steady-state phases of the DVM are independent of the initial configurations. However, as observed in Ref.~\cite{xue2023machine} for the VM, in the DVM also, phase boundaries between the two neighboring phases, the cross-sea phase and the microphase, are not always distinct. For a finite system size, bistability between these two neighboring phases can be observed, as shown in  Fig.~\ref{stabilityfig2}. At low noise ($\eta=0.1$), the system exhibits a homogeneous high-density ordered phase, where particles move mainly in a uniform direction. With a small increase of the noise $(\eta=0.2)$, the cross-sea phase starts to appear, and at high noise amplitude ($\eta=0.45$), the system transitions to a disordered phase, where particles show random orientations. Interestingly, in the intermediate noise amplitudes ($\sim \eta=0.28-0.35$), the system exhibits bistability where both cross-sea and microphase features can be found depending upon the initial conditions. In the VM, bistability was observed between the ordered phase and cross-sea phase and also between the microphase and disordered phase~\cite{xue2023machine}. This bistability arises due to the fluctuation at relatively higher noise amplitudes and finite system size. We therefore expect the system to evolve to a specific stable steady-state phase for a particular noise amplitude at the thermodynamic limit $(L \to \infty)$. The bistability between the cross-sea phase and the microphase as a function of noise amplitude is further demonstrated by a hysteresis loop in Fig.~\ref{stabilityfig3}. The hysteresis loop is obtained by plotting $\phi_{\rm cs}$ as a function of noise $\eta$ keeping the density $(\rho_0)$ and particle velocity $(v_0)$ constant. Where $\phi_{\rm cs}$ is either 1 or 0, corresponding to cross-sea and microphase, respectively. In the low noise limit, we evolve an initially disordered configuration into a steady state coexistence phase by slowly varying the noise amplitude by $(\Delta\eta=0.02)$. The cross-sea phase demonstrated by $\phi_{\rm cs}=1$ starts $\sim \eta=0.2$ and is retained up to a large noise (red curve). At $\sim \eta=0.35$, the cross-sea switches to the microphase and remains steady for up to $\sim \eta=0.4$. While reducing the noise from this high value (blue curve), the microphase bands $(\phi_{\rm cs}=0)$ sustain for a wide range of $\eta$ stretching significantly below the cross-sea to microphase transition point ($\sim \eta=0.35$). Finally, the microphase switches back to the cross-sea phase around ($\sim \eta=0.28$). As shown in Fig.~\ref{stabilityfig3}, both the transitions (at small and large $\eta$) between these two phases appear to be discontinuous.

\section{Impact of spatial anisotropy: steady-states for rectangular domain}
\label{appC}
\begin{figure*}
    \centering
    \includegraphics[width=\textwidth]{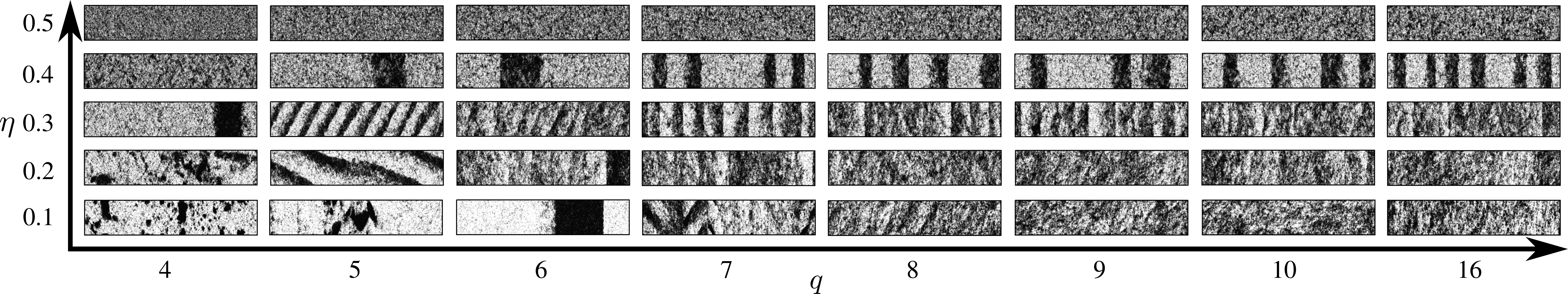}
    \caption{$\eta-q$ phase diagram of the DVM illustrated by snapshots on a rectangular domain ($L_x=800$, $L_y=100$). Parameters: $\rho_0=2$, $v_0=0.5$. As a function of $\eta$ and $q$, we observe five distinct self-organized patterns: cluster ($\eta=0.1$, $q=4$), macrophase separation ($\eta=0.3$, $q=4$), microphase separation ($\eta=0.4$, $q=7 \to 16$), ordered liquid ($\eta=0.2$, $q=7 \to 16$), and disordered gas ($\eta=0.5$, $q=4 \to 16$).}
    \label{appfig2}
\end{figure*}
In Fig.~\ref{appfig2}, we investigate how spatial anisotropy influences the non-equilibrium steady-state behavior of the DVM if we switch from a square domain to a rectangular domain by analyzing late-stage representative snapshots as a function of noise strength ($\eta$) and discretization parameter ($q$). When noise is low and $q$ is small, we observe the emergence of a locally ordered cluster phase similar to our finding for the square domain. However, when $q$ is small and fluctuations are pronounced, those cluster phases relax and transform into a larger, organized domain (see the snapshot for $q=4$ and $\eta=0.3$) akin to the Fig.~\ref{TDVM_L1024}. Conversely, when $\eta$ is small and $q$ is large, weak anisotropy facilitates the merger of the cluster phase into a larger, well-organized domain (see the snapshot for $q\geqslant8$ and $\eta=0.1$) which is observed at $q=10$ in the absence of spatial anisotropy (see Fig.~\ref{TDVM_L1024}). An increase in the fluctuation for small $q$ might exhibit multiple bands but those are connected by the periodic boundary conditions and should be considered a single band (see the snapshot for $q=5$ and $\eta=0.2, 0.3$). However, no cross-sea phase is observed with the rectangular geometry, which is seen in Ref~\cite{kursten2020dry} probably due to larger particle velocity ($v_0=1$). An increase in anisotropy $q$ also increases the no. of bands in the coexistence region at intermediate noise ($\eta=0.4$) because with $q$, the density fluctuation increases along with the magnetization fluctuation which prompts the breaking of large domains.

\section{$(\eta-\rho_0)$ phase diagram of the DVM}
\label{appD}
\begin{figure}
    \centering
    \includegraphics[width=\columnwidth]{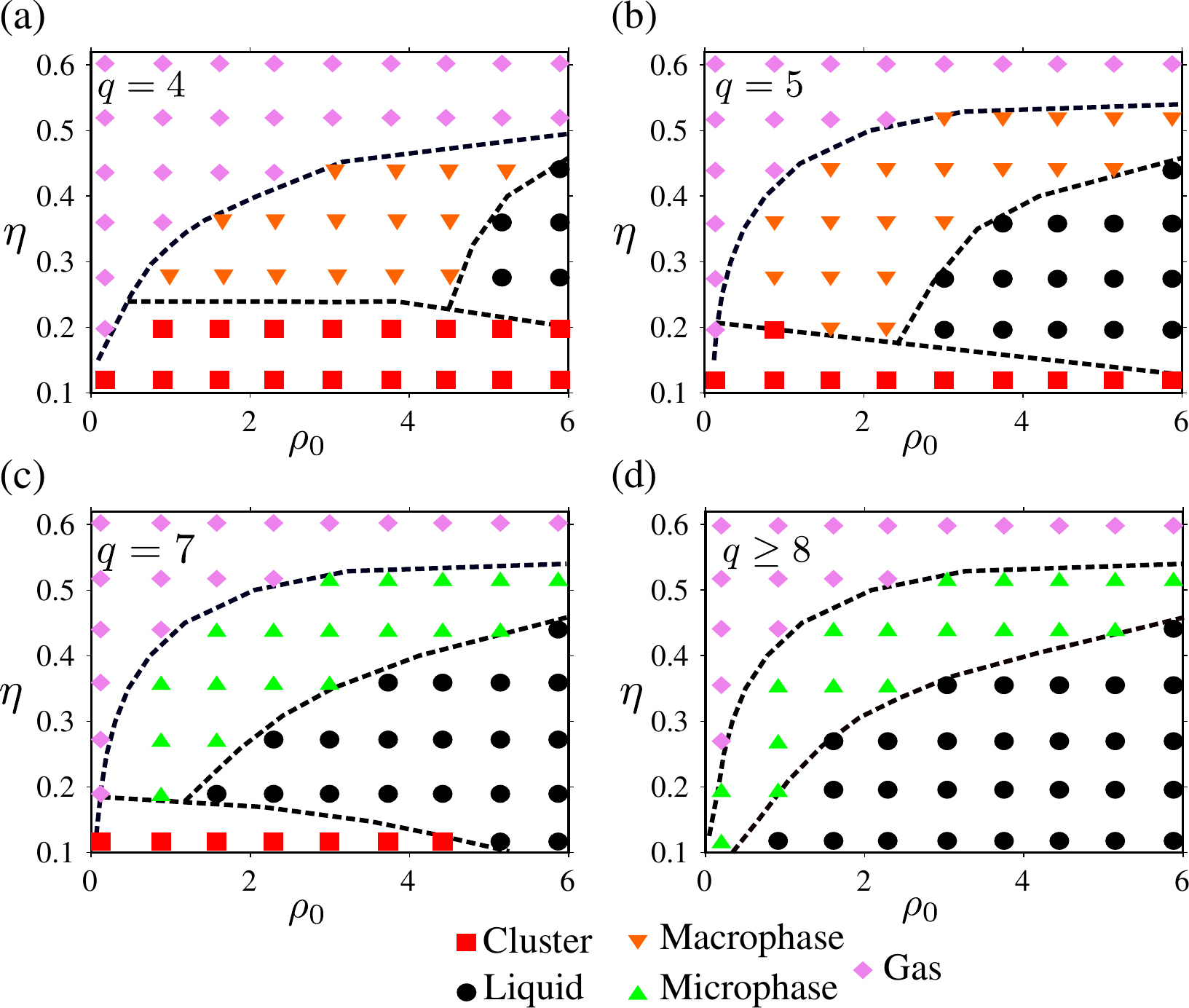}
    \caption{(Color online) Phase diagrams on $\eta-\rho_0$ plane computed on a rectangular domain of $800 \times 100$ for various $q$. A macrophase separation similar to AIM in (a) and (b) transforms to a Vicsek-like phase separation (microphase) in the coexistence region in (c) and (d) at intermediate densities. $v_0=0.5$.}
    \label{appfig3}
\end{figure}
In Fig.~\ref{appfig3}, we plot the noise-density ($\eta-\rho_0$) phase diagrams computed on a rectangular domain of $800 \times 100$ for various $q$. The clustering phase is very prominent for small $q$ values ($q=4$, 5) and exists for high densities. As $q$ is increased, the conventional ordered liquid phase appears at high densities and for $q \geqslant 8$, the cluster phase disappears and we notice the emergence of the typical $\eta-\rho_0$ diagram observed for Vicsek-like systems [Fig.~\ref{appfig3}(d)]. These Vicsek-like phases are characterized by a more global alignment of the particles, leading to coherent motion and the absence of distinct clusters or bands. In this regime, the behavior of the system is predominantly governed by the alignment interactions between the particles rather than the specific value of $q$. It shows clear boundaries between different phases based on varying values of $q$ and $\eta$. At small $q$ and intermediate densities, a macrophase separation similar to AIM is observed. However, as $q$ exceeds a threshold (e.g., $q \geqslant 8$), the system transitions to a Vicsek-like phase separation in the coexistence region. 

\section{Direction of global order in the coexistence region}
\label{appF}
\begin{figure}
    \centering
    \includegraphics[width=\columnwidth]{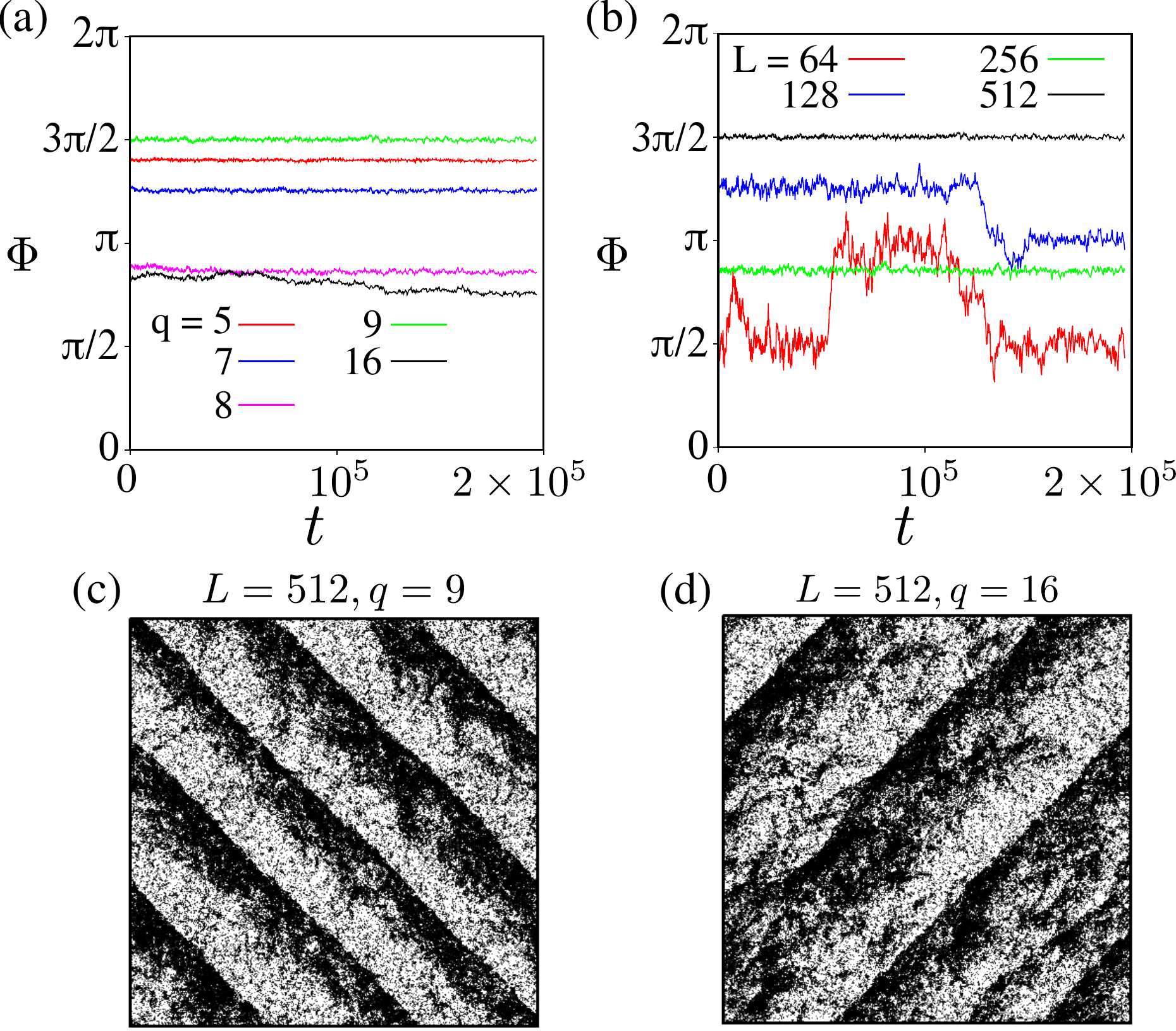}
    \caption{(Color online) Time evolution of the direction of global order $\Phi$ in the coexistence region for (a) $L=512$ and various $q$ and (b) $q=9$ and several values of $L$. (c--d) Snapshots showing multiple bands moving parallelly constituting a microphase-separated coexistence region for $q=9$ and $q=16$, respectively. System size $L=512$. Parameters: $\eta=0.3$, $\rho_0=1.5$, $v_0=0.5$.}
    \label{appfig8}
\end{figure}
Here, we analyze the pinned property of the system in the coexistence region and directly compare the outcome with the steady-state snapshots. In Fig.~\ref{appfig8}, the time series of the orientation of global order $\Phi$ is shown for fixed $L=512$ and various $q$ [Fig.~\ref{appfig8}(a)] and for fixed $q=9$ and several $L$ [Fig.~\ref{appfig8}(b)]. When the system size is fixed, we observe pinning to unpinning transition with $q$ whereas for fixed $q$, the reverse transition happens as the system size increases. Both observations are similar to the observations made regarding the time evolution of $\Phi$ in the DVM polar ordered phase [Fig.~\ref{figGOP}]. The snapshots shown in Fig.~\ref{appfig8}(c--d) without any spatial anisotropy exhibit a coexistence region with multiple bands signifying a microphase-separated region. Comparing the snapshots with the time evolution of $\Phi$ we notice that both the unpinning behavior of the DVM for $q=16$ and the pinning behavior for $q=9$ shows flocking with multiple parallelly moving bands. 

\section{Nature of the DVM ordered phase}
\label{appE}
\begin{figure}
    \centering
    \includegraphics[width=\columnwidth]{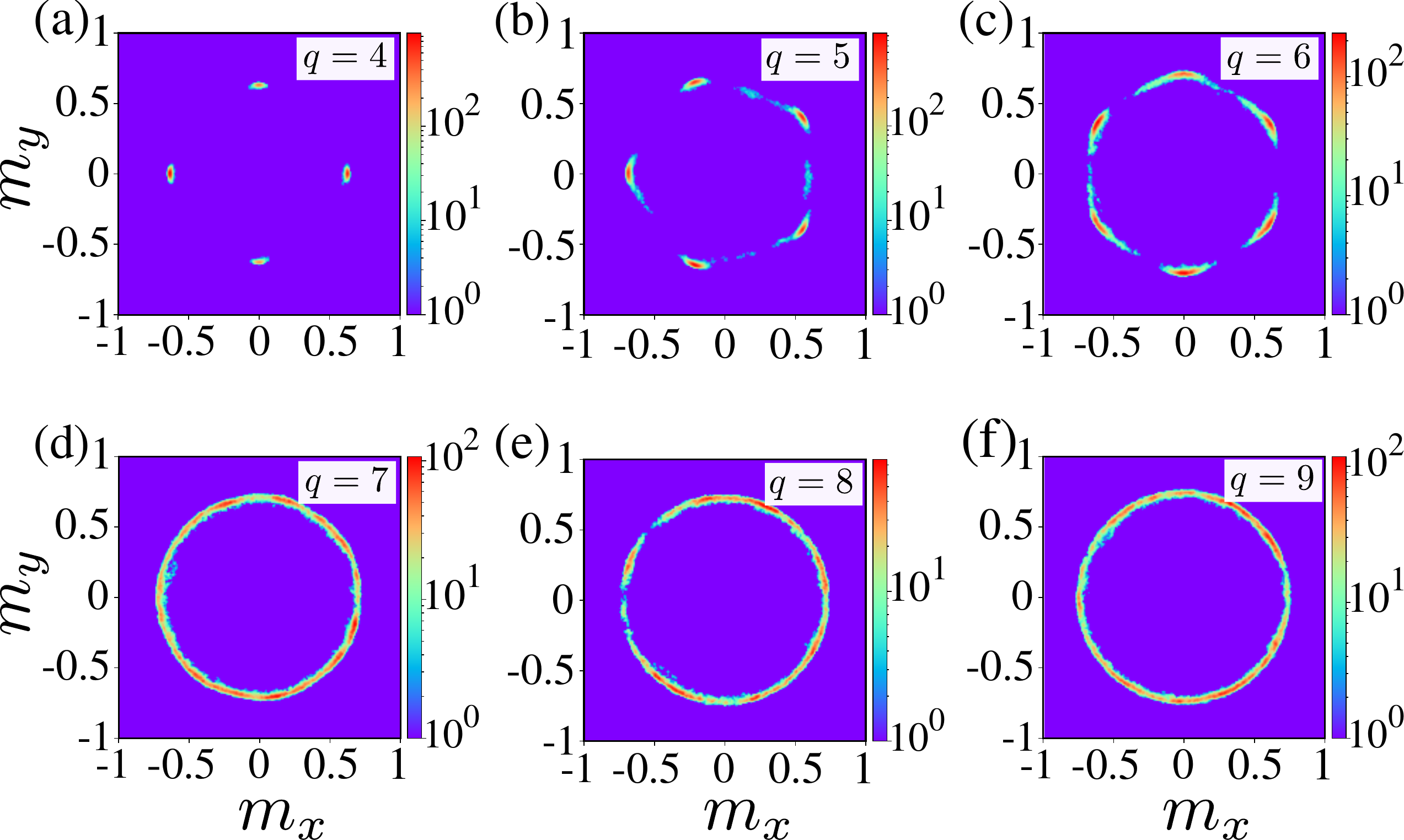}
    \caption{(Color online) Zero-activity limit of the order parameter distributions of the DVM with varying $q$ in the liquid phase. Parameters: $L=100, \eta=0.3$, and $\rho_0=6$. Distinct isolated spots in (a), (b), and (c) indicate LRO while ring-like distributions in (d), (e), and (f) are characteristic of the QLRO phase.}
    \label{appfig5}
\end{figure}

\begin{figure}
    \centering
    \includegraphics[width=\columnwidth]{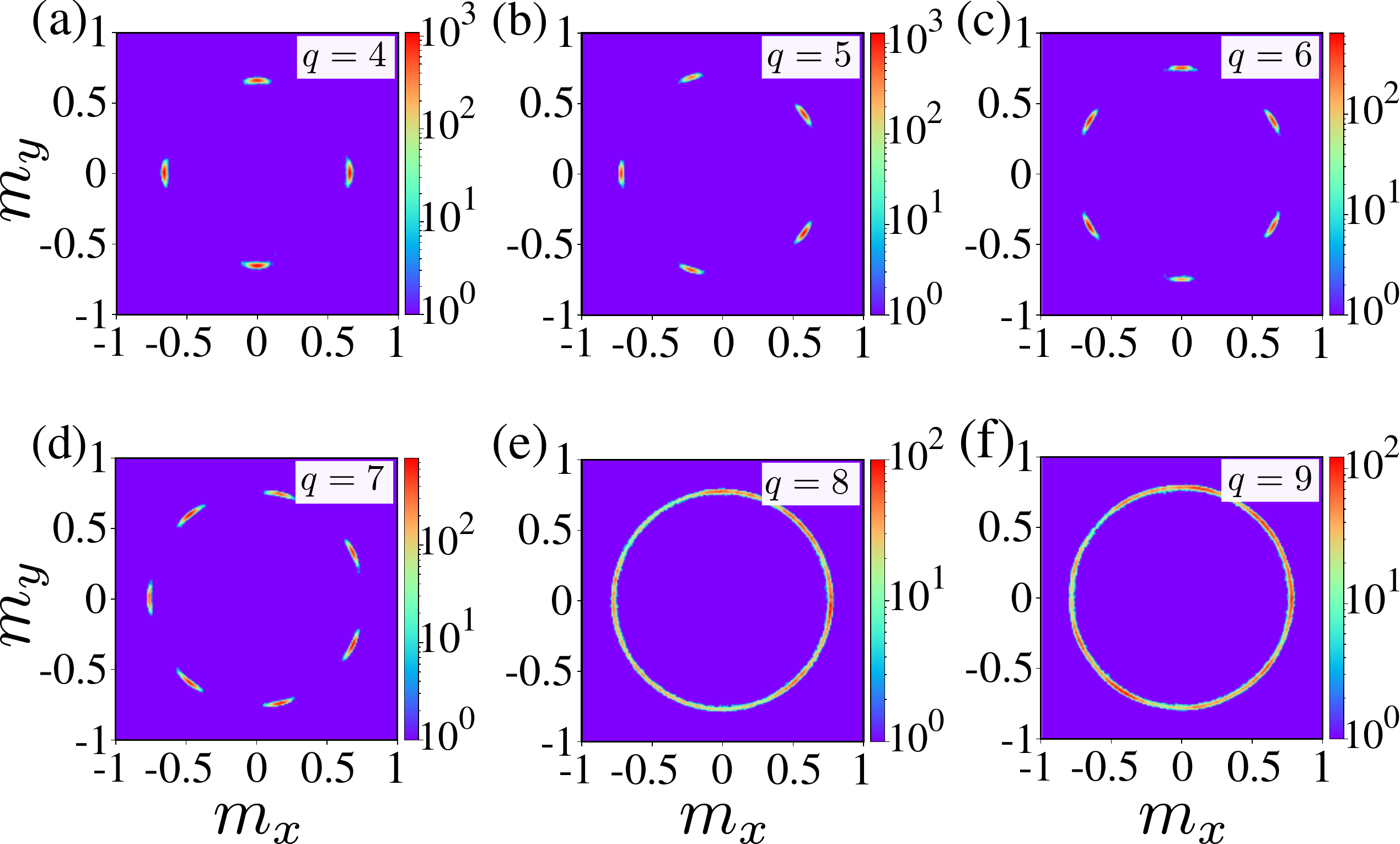}
    \caption{(Color online) Order parameter distributions of the DVM at constant activity $v_0=0.5$ with varying $q$ in the liquid phase. Parameters: $L=100, \eta=0.3$, and $\rho_0=6$.}
    \label{appfig6}
\end{figure}
In Fig.~\ref{appfig5}, we show the order parameter ({\bf m}) distribution on the $(m_x,m_y)$ plane for $v_0=0$. At this limit, particles can no longer move but modify their orientation according to Eq.~\eqref{sigma} and the $q$-state DVM reduces to the two-dimensional $q$-state clock model which shows two distinct phase transitions, one from disordered to QLRO phase at a higher temperature and the other from QLRO to LRO phase at a lower temperature for $q \geqslant 5$ \cite{clockmodel2018}. At large $q$, the LRO phase gradually starts to vanish, and for the two-dimensional XY model ($q \to \infty$), only one phase transition occurs (Kosterlitz-Thouless phase transition) from the disordered phase to the QLRO phase. As shown in Fig.~\ref{appfig5}, at the zero activity limit, the order parameter distribution shows $q$ distinct isolated spots (signifying LRO) for small $q$ but as $q$ increases ($q\geqslant 7$), ring-like distributions characteristic of the QLRO phase appears. These ring-like distributions become more pronounced as the system size is increased for large $q$. In contrast, for $v_0>0$ (see Fig.~\ref{appfig6}), the order parameter distribution exhibits LRO through isolated points of phase ordering as activity facilitates the broken symmetry phase. For $q>7$, a comprehensible LRO phase is observed at a large length scale limit as shown in Fig.~\ref{op_distro}. 
\begin{figure}
    \centering
    \includegraphics[width=\columnwidth]{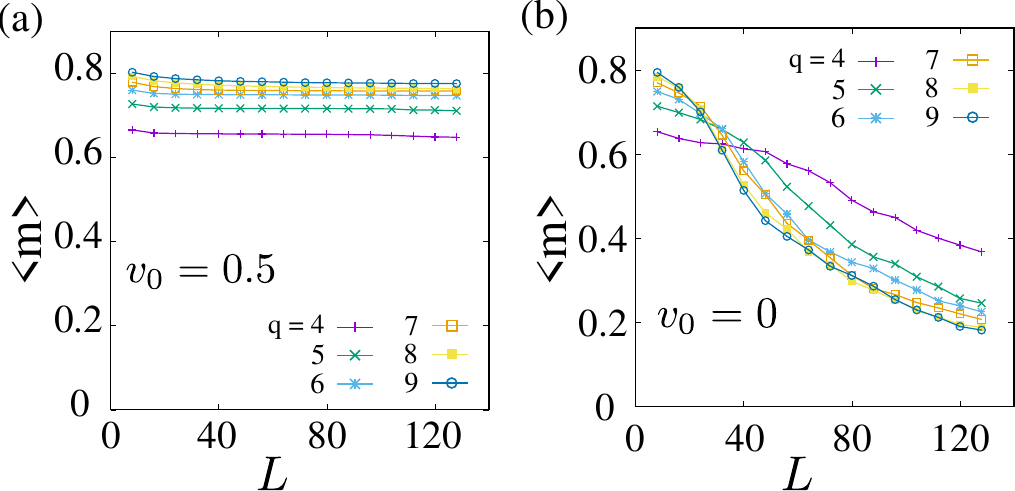}
    \caption{(Color online) Order parameter $\langle m \rangle$ versus system size $L$ for varying $q$ in the ordered liquid phase. (a) $\langle m \rangle$ is independent of $L$ for $v_0 = 0.5$ signifying LRO. (b) $\langle m \rangle$ decays algebraically with $L$ signifying QLRO in the zero-activity limit. Parameters: $\eta=0.3$, $\rho_0=6$.}
    \label{appfig7}
\end{figure}

To understand better the nature of ordering of the DVM liquid phase, we show the order parameter $\langle m \rangle$ against increasing system size $L$ for several $q$ in Fig.~\ref{appfig7}. The data presented are averaged over time and several initial configurations. We note that, $\langle m \rangle$ remains independent of the system size $L$ (actually, $m$ scales with $L$ as $m \sim L^{-\lambda}$, decays much slower than a power law) for all $q$ for $v_0 = 0.5$ [Fig.~\ref{appfig7}(a)] signifying LRO. As a result, the liquid phase of the constant-speed DVM is LRO and the direction of the order parameter exhibits a pinned behavior. For $v_0 = 0$, shown in Fig.~\ref{appfig7}(b), however, $\langle m \rangle$ is expected to algebraically decay to zero for $L\to \infty$ and this effect is more pronounced for larger $q$ because, for large $q$, the $v_0=0$ DVM approaches the two-dimensional XY model where the low-temperature phase is QLRO.

\bibliography{TDVM_biblio}
\end{document}